\documentclass[12pt]{article}
\pdfoutput=1
\usepackage{amsmath,amsfonts,graphicx,color,bbm,tikz,bm,setspace}
\usepackage{comment}
\usepackage[nosort]{cite}
\usepackage{subfigure}
\usepackage{qcircuit}
\usepackage{multirow}
\usetikzlibrary{calc,positioning}
\usetikzlibrary{patterns,arrows,decorations.pathreplacing}
\usepackage{caption}
\usepackage{ulem}
\tikzset{>=stealth}
\usepackage{lipsum}
\usepackage{tabularx}
\usepackage{fullpage}
\usepackage{hyperref}
\usepackage{bbold}
\usepackage{amsthm}
\theoremstyle{definition}
\newtheorem{definition}{Definition}[section]
\newtheorem{proposition}{Proposition}[section]
\newtheorem{remark}{Remark}[section]

\usepackage{listings}
\usepackage{tcolorbox}

\newcommand{\mathsym}[1]{{}}
\newcommand{\unicode}[1]{{}}

\makeatletter\@addtoreset{equation}{section}\makeatother

\newcommand{\be}{\begin{equation}}
\newcommand{\ee}{\end{equation}}
\def\beq{\begin{equation}}
\def\eeq{\end{equation}}
\newcommand{\bea}{\begin{eqnarray}}
\newcommand{\eea}{\end{eqnarray}}

\def\nn{\nonumber}

\renewcommand{\title}[1]{\vbox{\center\LARGE{#1}}\vspace{3mm}}
\renewcommand{\author}[1]{\vbox{\center{#1}}\vspace{3mm}}

\newcommand{\email}[1]{\vbox{\center\tt#1}\vspace{3mm}}

\begin{document}

\begin{center}
{\large {\bf Unitary tetrahedron quantum gates}}

\author{Vivek Kumar Singh,$^a$ Akash Sinha,$^b$ Pramod Padmanabhan,$^b$ Vladimir Korepin$^{c}$}

{$^a${\it Center for Quantum and Topological Systems (CQTS), NYUAD Research Institute,\\ New York University Abu Dhabi, PO Box 129188, Abu Dhabi, UAE}}
\vskip0.1cm
{$^b${\it School of Basic Sciences,\\ Indian Institute of Technology, Bhubaneswar, 752050, India}}
\vskip0.1cm
{$^c${\it C. N. Yang Institute for Theoretical Physics, \\ Stony Brook University, New York 11794, USA}}

\email{vks2024@nyu.edu, akash26121999@gmail.com, pramod23phys@gmail.com, vladimir.korepin@stonybrook.edu}

\vskip 0.5cm 

\end{center}


\abstract{
\noindent 
Quantum simulations of many-body systems using 2-qubit Yang-Baxter gates offer a benchmark for quantum hardware. This can be extended to the higher dimensional case with $n$-qubit generalisations of Yang-Baxter gates called $n$-simplex operators. Such multi-qubit gates potentially lead to shallower and more efficient quantum circuits as well. Finding them amounts to identifying unitary solutions of the $n$-simplex equations, the building blocks of higher dimensional integrable systems. These are a set of highly non-linear and over determined system of equations making it notoriously hard to solve even when the local Hilbert spaces are spanned by qubits. We systematically  overcome this for higher simplex operators constructed using two methods: from Clifford algebras and by lifting Yang-Baxter operators. The $n=3$ or the tetrahedron case is analyzed in detail. For the qubit case our methods produce 13 inequivalent families of unitary tetrahedron operators. 12 of these families are obtained by appending the 5 unitary families of 4 by 4 constant Yang-Baxter operators of Dye-Hietarinta, with a single qubit operator. As applications, universal sets of single, two and three qubit gates are realized using such unitary tetrahedron operators. The ideas presented in this work can be naturally extended to the higher simplex cases. 
}


\section{Introduction}
\label{sec:Introduction}
The solutions of the Yang-Baxter equation \cite{YangCN1967,BAXTER1972193}, known either as $R$-matrices, Yang-Baxter operators or 2-simplex operators, first appeared in the context of exactly solvable two dimensional statistical mechanical models associated to one dimensional quantum spin chains \cite{Baxter1982ExactlySM}. This laid the framework for analytically solving the spectrum and correlation functions of several quantum systems in statistical mechanics and condensed matter physics, including the Heisenberg and Hubbard models, using the method of algebraic Bethe ansatz \cite{Korepin1993QuantumIS,takhtadzhyan1979,Takhtadzhan_1979,slavnov2019algebraicbetheansatz,Franchini2016AnIT,Retore2021IntroductionTC,Arutyunov2019ElementsOC,LevkovichMaslyuk2016TheBA,Batchelor2015YangBaxterIM,Doikou2009IntroductionTQ,Nepomechie1998ASC,eckle2019models}. In this setting the Yang-Baxter operator appears as a function of spectral parameter(s), which for particular values reduces to braid group generators that are devoid of these parameters. The latter furnish representations of the braid group and solve the constant Yang-Baxter equation that are precisely the braid relations. This helps study knot and link invariants \cite{Kauffman1991,Turaev1988TheYE}. This perspective leads to a conjectured connection between topological entanglement and quantum entanglement \cite{kauffman2002quantum,kauffman2003entanglement,Aravind1997} where a correspondence between entangled states in multi-qubit systems and knots and links is established \cite{Quinta_2018}.

A more modern application of the Yang-Baxter operator is to use their unitary versions as quantum gates that can realize universal quantum computation \cite{kauffman2004braiding,zhang2005universal}. Such gates are the building blocks of a model of integrable quantum computation that are speculated to be prone to much lesser noise than other paradigms \cite{zhang2013integrable,lloyd2014universal,Zhang2021QuantumCU}. A closely related application is the use of Yang-Baxter gates in integrable quantum circuits \cite{Miao2023IntegrableQC, Miao2022TheFB, Claeys:2021cdy, aleiner2021bethe, Vernier:2024cnw, Zhang_2024, zhang2024geometricrepresentationsbraidyangbaxter}. These gates depend on a spectral parameter and hence they solve either the additive or non-additive form of the spectral parameter dependent Yang-Baxter equation. These gates are used in obtaining the Hamiltonian evolutions of a chain of qubits by Trotterization resulting in quantum circuits describing a space-time discretized system or in other words, a simulation of a many-body system. Such circuits are hard to obtain and solve for a generic quantum system. However this is greatly simplified when the system is integrable \cite{Vanicat2017IntegrableTL} as the associated quantum circuits, even if deeper to better approximate the evolution, is nevertheless better protected from noise due to the integrability. Thus integrable systems described by $R$-matrices help to develop quantum algorithms used in the simulation of quantum many-body systems, that can eventually be used to benchmark the associated quantum hardware. Naturally this process also benefits integrable systems as it helps to study the algebraic Bethe ansatz on a quantum computer \cite{aleiner2021bethe,Sopena_2022,VanDyke2021PreparingBA}. The hope here is to explore higher excited states that are computationally costly on a classical computer. In a related interesting development integrable systems {\it via} $R$-matrices are shown to also play a role in describing systems out of equilibrium through generalized Gibbs ensembles \cite{Vidmar2016GeneralizedGE,Essler_2016} and generalized hydrodynamics \cite{PhysRevX.6.041065,PhysRevLett.117.207201,RevModPhys.93.025003,Alba_2021}. The integrable quantum circuits for such systems have gained much attention as well \cite{sa2021integrable,su2022integrable}. In this case the Yang-Baxter operators approximate continuous-time integrable Lindblad superoperators \cite{10.21468/SciPostPhys.8.3.044,PhysRevLett.126.240403}. Thus these studies open another set of applications for the Yang-Baxter operators beyond their intended ones. 

The $R$-matrix plays a major role, as is evident in all of the stated applications of the Yang-Baxter equation. Thus finding these solutions has been in focus for quite some time now\footnote{There are different methods to produce such solutions, including numerical techniques, set-theoretic solutions \cite{Etingof1998SettheoreticalST}, and algebraic methods. The resulting operators come with or without spectral parameters, which can be either additive or non-additive when they are present. The solutions include both invertible and non-invertible operators. When they are invertible they can be either unitary or non-unitary.}. In this work we lay the framework to extend all of the above applications to higher dimensions ($>2$) by producing unitary versions of $n$-qubit generalizations of the 2-qubit Yang-Baxter operator. These operators, called the $n$-simplex operators, solve the $n$-simplex equations \cite{CarterSaito,Frenkel1991SimplexEA,MAILLET1989221} which are proposed to give rise to higher dimensional integrable systems \cite{Zamolodchikov1980TetrahedronOriginal,Baxter1983OnZS,Maillet1989IntegrabilityFM,Bazhanov1982ConditionsOC,Bazhanov1992NewSL,Bazhanov1993StartriangleRF,Stroganov1997TetrahedronEA,Mangazeev2013AnI3,Sergeev1995TheVF,Talalaev2015ZamolodchikovTE,Khachatryan2015IntegrabilityIT,Talalaev2021TetrahedronEA,kuniba2022quantum}, analogous to the role played by the Yang-Baxter equation in two dimensional integrable systems. The most well-studied among these are the tetrahedron or 3-simplex equations. These equations first appeared in the context of the scattering of straight lines in $2+1$-dimensions \cite{Zamolodchikov1980TetrahedronOriginal,Zamolodchikov1981TetrahedronEA}. Since then several solutions have been formulated using various methods\cite{Gavrylenko2020SolutionOT,inoue2024tetrahedron,Kuniba2012TetrahedronA3,sun2022cluster,inoue2023quantum,inoue2024solutions,bardakov2022settheoretical, Hietarinta_1993,Hietarinta_1997,iwao2024tetrahedronequationschurfunctions,Kuniba2015TetrahedronEA,Kuniba2022NewST,Kuniba2015TetrahedronEA-2}.

For our purposes we will use the solutions obtained using Clifford algebras \cite{padmanabhan2024solving} and by lifting Yang-Baxter operators to the 3-qubit case \cite{Hietarinta_1993}. The main result of our work is the classification of the unitary versions of the tetrahedron operators constructed using these methods. We find 13 inequivalent families of such operators. The solutions are designed in a way that they find applicability in quantum many-body physics, as quantum gates and in the construction of integrable quantum circuits to simulate higher dimensional systems. In the latter the holy grail will be to solve the 3D Ising model with methods similar to Bethe ansatz for the tetrahedron equation, with some studies already taken in this direction using tetrahedron equations \cite{talalaev2018integrablestructure3dising}. However this is beyond the scope of the current work, but we hope to have taken the first steps in this direction. As quantum gates, tetrahedron operators provide multi-qubit gates that can potentially realize universal quantum computation in a more efficient manner with shallower quantum circuits. We show how several standard one, two and three qubit gates are realized using tetrahedron solutions.

Our results are organized as follows. We begin with a review of the different types of tetrahedron equations \cite{Hietarinta1997TheET} and recall their discrete and continuous symmetries in Sec. \ref{sec:preliminaries}. This section also includes a discussion on the various properties of unitary matrices and methods to unitarize a given invertible matrix. The first set of unitary tetrahedron operators using Clifford algebras is constructed in Sec. \ref{sec:unitaryCliffordSolutions}. This accounts for just one of the unitary families in the final tally. The remaining unitary solutions constructed from Yang-Baxter operators is the subject of Sec. \ref{sec:TetraFromYBsolutions}. After a discussion on the general construction (representation independent) we study two sets of tetrahedron operators : the first set (Sec.\ref{subsec:TetrafromHietarinta}) constructed from the constant 4 by 4 Yang-Baxter classification of Hietarinta \cite{Hietarinta_1992} and the set second (Sec. \ref{subsec:TetrafromClifford}) obtained by appending single qubit operators to Yang-Baxter solutions constructed using Clifford algebras. However the latter are gauge equivalent to the former and hence do not produce new unitary families. As a result we find that the former produces 12 unitary families (obtained as an extension of the 5 unitary families of Yang-Baxter operators \cite{dye2003unitarysolutionsyangbaxterequation}) of tetrahedron operators as summarized in Sec. \ref{sec:summaryTetra}. We close with a discussion on universal quantum gates sets using tetrahedron operators in Sec. \ref{sec:discussion}. 

Three appendices are included to fill in some details and to include some additional content that are outside the theme of the main text but nevertheless related. In Appendix \ref{app:lower2higher} we provide more examples of how to construct higher simplex operators by tensoring lower simplex operators with operators acting on a single local Hilbert space. This can be viewed as a generalization of the method provided in Hietarinta's paper \cite{Hietarinta_1993} to obtain tetrahedron operators from Yang-Baxter operators. Appendix \ref{app:3from2} explores the generalized tetrahedron equations satisfied by other possible words made out of three operators $A$, $B$ and $C$. This is a technical appendix included to complete the analysis of Sec. \ref{subsec:TetrafromClifford}. The third and final appendix \ref{app:BaxterizationT} considers `Baxterization' of tetrahedron operators, that is the process of introducing spectral parameters into constant tetrahedron operators.

\section{Preliminaries }
\label{sec:preliminaries}
The initial formulation of the tetrahedron equation was in terms of the scattering matrix elements generalizing the Yang-Baxter equation that describes the scattering of point particles in $1+1$-dimensions \cite{Zamolodchikov1979FactorizedSI,YangCN1967,BAXTER1972193}. The ambiguity in labelling the scattering process in two dimensions leads to alternative forms for the tetrahedron equation \cite{Hietarinta_1994,Hietarinta1997TheET,MAILLET1989221,Baez1995HigherDA}. In particular the vertex form reads :
\begin{equation}\label{eq:tetraVertexForm}
    R_{123}R_{145}R_{246}R_{356} = R_{356}R_{246}R_{145}R_{123}.
\end{equation}
This is the constant form (independent of spectral parameters) of the tetrahedron (3-simplex) equation. This is a constraint on the scattering of particles formed at the intersections of the four straight lines. On the other hand the edge form (also known as the Frenkel-Moore equation \cite{Frenkel1991SimplexEA}) of the tetrahedron equation labels the line segments of the scattering strings :
\begin{equation}\label{eq:tetraEdgeForm}
  R_{123}R_{124}R_{134}R_{234} = R_{234}R_{134}R_{124}R_{123}.  
\end{equation}
The tetrahedron operators satisfy certain symmetry conditions \cite{Hietarinta_1993}. The discrete symmetries include :
\begin{eqnarray}
    R_{ijk}^{lmn} & \rightarrow & R_{lmn}^{ijk} , \nonumber \\
    R_{ijk}^{lmn} & \rightarrow & R_{kji}^{nml} , \nonumber \\
    R_{ijk}^{lmn} & \rightarrow & R_{i+s,j+s,k+s}^{l+s, m+s, n+s}~~s>0. 
\end{eqnarray}
All indices are taken mod $dim~V=N$. The first equation corresponds to simple matrix transposition. It is easily verified that for a given solution, its inverse also form solutions. Furthermore an equivalence class of solutions is obtained {\it via} a local gauge principle. For every tetrahedron operator $R$, the set obtained by conjugating $R$ with a local similarity transform
\begin{equation}\label{eq:localEquiv}
   \mathcal{Q}R\mathcal{Q}^{-1}\equiv\kappa~\left(Q\otimes Q\otimes Q\right)~R~\left(Q^{-1}\otimes Q^{-1}\otimes Q^{-1}\right), 
\end{equation}
also satisfies the tetrahedron equation. Here $Q$ is an invertible operator and $\kappa$ is a complex constant. 

While these symmetry relations produce inequivalent classes of tetrahedron operators that include both invertible and non-invertible operators, the unitary solutions amongst them are harder to obtain. The reason for this subtlety is due to the fact that $\mathcal{Q}R\mathcal{Q}^{-1}$ may be unitary even if $R$ is not unitary. Thus to obtain unitary tetrahedron operators we need to study the equivalence classes of invertible tetrahedron operators generated by the local gauge action \eqref{eq:localEquiv}. This amounts to finding the $Q$ matrices in terms of the parameters of the invertible tetrahedron operators. 

In this work we assume that the tetrahedron operators $R_{ijk}$ act on finite dimensional vector spaces $V$ indexed by $i$, $j$ and $k$. The resulting operators are matrices acting on $V\otimes V\otimes V$. In other words we find matrix representations of the tetrahedron operators. So we recall some basic facts about unitary matrices :
\begin{enumerate}
    \item A matrix $M$ is unitary if $M^{-1} = M^\dag$, where the $\dag$\footnote{The literature also uses $*$ instead of $\dag$.} symbol stands for complex conjugation and transpose. 
    \item $(M^\dag)^{-1}=(M^{-1})^\dag$ if $M$ is invertible.
    \item The eigenvalues $\lambda$ of a unitary matrix satisfy $|\lambda|=1$.
    \item If $M$ is a unitary matrix and if $\alpha\in\mathbb{C}$, then $\alpha M$ is unitary iff $|\alpha|=1$.
\end{enumerate}
If the dimension of $V$ is $N$, then the vertex form of the tetrahedron equation \eqref{eq:tetraVertexForm} contains $N^{12}$ equations in $N^6$ variables implying that this is an highly overdetermined system. Our focus in this paper will be on qubit representations ($N=2$) or $V=\mathbb{C}^2$. This makes the tetrahedron operators 8 dimensional matrices. The task of identifying unitary versions of these matrices is analytically tedious and also numerically difficult. Thus we require ways to simplify this problem. To this end we will review the arguments used in \cite{dye2003unitarysolutionsyangbaxterequation} for classifying the constant 4 by 4 unitary Yang-Baxter solutions. We begin with the following proposition.
\begin{proposition}
    {\it Consider the operator $\mathcal{Q}R\mathcal{Q}^{-1}$. Then the following statements are true :
    \begin{enumerate}
        \item $\mathcal{Q}R^{-1}\mathcal{Q}^{-1} = (\mathcal{Q}^\dag)^{-1}R^\dag \mathcal{Q}^\dag$ and this implies $\mathcal{Q}^\dag\mathcal{Q}R^{-1} = R^\dag \mathcal{Q}^\dag\mathcal{Q}$.
        \item $R^\dag$ and $R^{-1}$ have the same set of eigenvalues.
    \end{enumerate}}
\end{proposition}
The proof is straightforward as it follows from the definition. The interested reader is referred to \cite{dye2003unitarysolutionsyangbaxterequation} for details. Note that this results holds for any pair of matrices $R$ and $\mathcal{Q}$. In particular, it is applicable to the case when $R$ is a tetrahedron operator and $\mathcal{Q}=Q\otimes Q\otimes Q$ as well. This result suggests that an indirect way to find a family of unitary matrices $\mathcal{Q}R\mathcal{Q}^{-1}$ is to equate the eigenvalues of $R^\dag$ and $R^{-1}$. We build on this idea to further simplify this process. Following the arguments given in \cite{dye2003unitarysolutionsyangbaxterequation} our next result uses an arbitrary invertible $Q$ matrix,
\begin{equation}\label{eq:Qmatrix}
    Q = \begin{pmatrix}
        q_1 & q_2 \\ q_3 & q_4
    \end{pmatrix},~~q_1,q_2,q_3,q_4\in\mathbb{C}.
\end{equation}
We then have
\begin{proposition}
    {\it If $q_1\bar{q_2} + q_3\bar{q_4} = 0$, then $Q^\dag Q$ and $H=\mathcal{Q}^\dag\mathcal{Q}$ are both diagonal.}
\end{proposition}
The proof is once again easy following some simple computations. Nevertheless we outline a few steps. We have 
\begin{equation}
    Q^\dag Q = \begin{pmatrix}
        \bar{q_1}q_1 + \bar{q_3}q_3 & \bar{q_1}q_2 + \bar{q_3}q_4 \\ 
        q_1\bar{q_2} + q_3\bar{q_4} & \bar{q_2}q_2 + \bar{q_4}q_4
    \end{pmatrix} \equiv \begin{pmatrix}
        x & \bar{z} \\ z & y
    \end{pmatrix}.
\end{equation}
This is diagonal when $z=0$ or when $q_1\bar{q_2} + q_3\bar{q_4} = 0$. The diagonal elements of the matrix $H=\mathcal{Q}^\dag\mathcal{Q}$ are functions of just $x=\bar{q_1}q_1 + \bar{q_3}q_3$ and $y=\bar{q_2}q_2 + \bar{q_4}q_4$. And all the off-diagonal elements have a factor of either $z$ or $\bar{z}$. Thus when $z=0$
$H$ is diagonal finishing the proof.
\begin{definition}
    {\it Consider the operator $D = HR^{-1} - R^\dag H$. This operator  measures the deviation from unitarity of the operator $R$, in the sense that when $D=0$, $R$ is unitary.}
\end{definition}
We will compute the matrix $D$ in each case to determine the unitary family of tetrahedron operators. The following two propositions will aid us in this process.
\begin{proposition}
   {\it For invertible $Q$, $H_{ii}\neq 0$, $\forall i$.}
\end{proposition}
This is easy to see as if $H_{ii}=0$, then either $x$ or $y$ is zero (recall that the diagonal elements of $H$ are functions of $x$ and $y$). If $x=0$ then $|q_1|^2+|q_3|^2=0$ implying both $q_1$ and $q_3$ are 0. This means $Q$ is non-invertible contradicting the assumption that $Q$ is invertible. A similar argument holds when $y=0$.
\begin{proposition}
  {\it Let $Q$ be invertible. If $H_{ij}=0$ for some $i,j$, then $H_{ij}=0$ $\forall~i,j$.}
\end{proposition}
Recall that every off-diagonal element of $H$, $H_{ij}$ when $i\neq j$, is a function of $z$ or/and $\bar{z}$ apart from either $x$ or $y$ or both. If one of the off-diagonal entries is 0 and this is a result of $z$ being 0, then all the other off-diagonal entries are also zero, proving the proposition. On the other hand if one of the off-diagonal entries is zero as a result of either $x$ or $y$ being zero, then we find that either $q_1$ and $q_3$ or $q_2$ and $q_4$ is zero, contradicting the assumption that $Q$ is invertible. Thus $H_{ij}=0$, only when $z=0$ and when this is true all the off-diagonal entries are 0, thus proving the assertion.

These two propositions essentially simplify the task of computing $D$ as they specify the restrictions on the matrix $H$ for invertible $Q$. As we shall see this helps in finding the conditions when $D=0$. These considerations lay the framework for computing unitary tetrahedron operators. Our next task is to construct the tetrahedron operators. We use two methods for this.
\begin{enumerate}
    \item The first set of solutions is constructed using two operators $A$ and $B$, that anticommute with each other and are realized using Clifford algebras. This method was first presented in \cite{padmanabhan2024solving} and so we borrow the solutions from there. These solutions solve both the vertex and the edge forms of the tetrahedron equation.
    \item The second set of solutions are constructed from Yang-Baxter operators. The index structure of the vertex form of the tetrahedron equation \eqref{eq:tetraVertexForm} suggests that such solutions exist. For example ignoring indices 3, 5 and 6 reduces the vertex tetrahedron equation to a constant Yang-Baxter equation in indices 1, 2 and 4. In the same manner removing indices 1, 2 and 3 gives us a Yang-Baxter or 2-simplex equation in 4, 5 and 6. We will construct two sets of such solutions.
    \begin{enumerate}
    \item The first set is constructed by tensoring a Yang-Baxter solution with a third operator. The technique shown here is adapted from a similar solution found in \cite{Hietarinta_1993}. The solutions are then written down using the 4 by 4 constant Yang-Baxter solutions of Hietarinta \cite{Hietarinta_1992}. The generalizations of this method to the higher simplex case is briefly discussed in Appendix \ref{app:3from2}. 
    \item The second set of solutions are special cases of the first set. 
    Here the Yang-Baxter operators are constructed using Clifford algebras \cite{padmanabhan2024solving} with a third operator $C$ (acting on a third space), that either anti-commutes or commutes with the $A$ and $B$ operators used to construct the Yang-Baxter solutions.
    \end{enumerate}
\end{enumerate}
In each of the above cases we identify the unitary solutions following the construction of the tetrahedron operators.

\section{Clifford tetrahedron solutions} 
\label{sec:unitaryCliffordSolutions}
The Clifford tetrahedron operators are based on two anticommuting operators $A$ and $B$ :
$$ AB = -BA.$$ The possible choices for $A$ and $B$ are exhausted by realizing them using Clifford algebras of an arbitrary order. Using them we can construct two kinds of tetrahedron operators, 
\begin{equation}\label{eq:tetraOP}
    R_{ijk} = \alpha_0~B_iB_jB_k + \alpha_1~A_iA_jB_k + \alpha_2~A_iB_jA_k + \alpha_3~B_iA_jA_k, 
\end{equation}
and 
\begin{equation}\label{eq:tetraOPBA}
    R_{ijk} = \alpha_0~A_iA_jA_k + \alpha_1~B_iB_jA_k + \alpha_2~B_iA_jB_k + \alpha_3~A_iB_jB_k. 
\end{equation}
According to the terminology used in \cite{padmanabhan2024solving}, the former is a linear combination of the $(0,3)$ and $(2,1)$ types, while the latter is a linear combination of the $(3,0)$ and $(1,2)$ types. The proof that these operators solve the tetrahedron equation is discussed in detail in \cite{padmanabhan2024solving}. These two operators can be equivalent to each other under the local gauge equivalence in \eqref{eq:localEquiv}. One way this can happen is if $A$ and $B$ are themselves equivalent by a similarity transform $Q$. Notice also that these solutions change according to the representation chosen for $A$ and $B$. In what follows we will represent them as matrices acting on local qubit Hilbert spaces. The generalizations to qudit Hilbert spaces will require a separate analysis which we defer to a future work. 

Two anticommuting operators $A$ and $B$ (both of them 2 by 2 matrices) can further satisfy one of three types of relations :
\begin{enumerate}
    \item $A^2=B^2=\mathbb{1}$. Either one or both of them can also square to $-\mathbb{1}$, but these can be obtained by scaling the operators with $\mathrm{i}=\sqrt{-1}$.
    \item Either one of them is nilpotent while the other squares to $\mathbb{1}$. For example we can have $A^2=0$ and $B^2=\mathbb{1}$. Here too one of the operators can square to $-\mathbb{1}$ by scaling with the complex factor $\mathrm{i}$. 
    \item In the third and final case both operators are projectors, $A^2=A$ and $B^2=B$. However in this case we also have $AB=BA=0$ or in other words $A$ and $B$ are orthogonal projectors. This case is special as the operators $A$ and $B$ both commute and anticommute with each other as they are orthogonal to one another. 
\end{enumerate}
The orthogonal projectors for $A$ and $B$ substituted into  \eqref{eq:tetraOP} and \eqref{eq:tetraOPBA} result in non-invertible tetrahedron operators. This beats our purpose of constructing unitary tetrahedron operators. Fortunately there is a way around this as we shall soon see in the following discussion of the three different cases. A remark is in order to determine the unitary families of solutions produced in each of these three cases. 
\begin{remark}
    The algebraic relations between the $A$ and $B$ operators, in each of the three cases, remain unchanged under conjugation by $Q$ ($A\rightarrow QAQ^{-1}$ and $B\rightarrow QBQ^{-1}$) as given by (\ref{eq:localEquiv}). From this we conclude that the local gauge equivalent solutions (\ref{eq:localEquiv}) do not change the conditions for each of the tetrahedron operators in the above three cases to be unitary. Thus it suffices to just study when the representatives \eqref{eq:tetraOP} and \eqref{eq:tetraOPBA} are unitary to determine the unitary families in each case.
\end{remark}  

\paragraph{Case 1 : $A^2=B^2=\mathbb{1}$ -}The operators $A$ and $B$ can be realized by Clifford generators which are Hermitian. Then for unitary tetrahedron operators the inverse of the operator in \eqref{eq:tetraOP} becomes 
\begin{equation}\label{eq:InverseTetraOP}
  R^{-1}_{ijk}=R^\dag_{ijk} = \Bar{\alpha}_0~B_iB_jB_k + \Bar{\alpha}_1~A_iA_jB_k + \Bar{\alpha}_2~A_iB_jA_k + \Bar{\alpha}_3~B_iA_jA_k.
\end{equation}
It is natural to assume that the inverse takes the same form as the original operator as the former has to solve the tetrahedron equation. Thus we are justified in looking for the unitary solutions within the same types of tetrahedron operators. We can make analogous statements for the tetrahedron operator in \eqref{eq:tetraOPBA}, but we take $A$ and $B$ to be equivalent by a similarity transform $Q$. So we only work with the operator in \eqref{eq:tetraOPBA}.

In a few steps, one can show that if the following conditions hold among the coefficients $\alpha$'s, the relation in (\ref{eq:InverseTetraOP}) is satisfied:
\begin{eqnarray}\label{eq: Z_2 sols}
\alpha_0 \bar{\alpha}_1-\alpha_3 \bar{\alpha}_2-\alpha_2 \bar{\alpha}_3+\alpha_1 \bar{\alpha}_0=0&,&~
-\alpha_3 \bar{\alpha}_1+\alpha_0 \bar{\alpha}_2-\alpha_1 \bar{\alpha}_3+\alpha_2 \bar{\alpha}_0=0,\nonumber\\
-\alpha_2 \bar{\alpha}_1-\alpha_1 \bar{\alpha}_2+\alpha_0 \bar{\alpha}_3+\alpha_3 \bar{\alpha}_0=0&,&~
\alpha_1 \bar{\alpha}_1+\alpha_2 \bar{\alpha}_2+\alpha_3 \bar{\alpha}_3+\alpha_0 \bar{\alpha}_0=1.
\end{eqnarray}
The parameters in \eqref{eq: Z_2 sols} are solved by setting $\alpha_j=r_je^{\mathrm{i}\theta_j}$. We find that the absolute values of the four complex numbers are on the surface of $S^3$ :
\begin{equation}
    \sum\limits_{j=0}^3~r_j^2 = 1,
\end{equation}
and they further satisfy 
\begin{eqnarray}
    r_0r_i\cos{(\theta_0-\theta_i)} = r_jr_k\cos{(\theta_j-\theta_k)}~;~i\neq j\neq k\in\{1,2,3\}.
\end{eqnarray}
Finding a general solution to this set of equations in complex variables is a hard task. We can obtain representation independent solutions in special situations. The conditions include setting two, three or four of the $r_j$'s equal to each other. Starting with the latter, we set $r_j=\frac{1}{2}, \forall j$ to obtain the solutions in Table \ref{tab:all4equal-Case1}
\begin{table}[h!]
\begin{center}
\begin{tabular}{|c|c|c|c|}
    \hline
   $\alpha_0$  & $\alpha_1$ & $\alpha_2$ & $\alpha_3$ \\
   \hline
   $e^{{\rm i}\psi} $ & $e^{{\rm i}\psi}$ & $e^{{\rm i}\phi}$ & $e^{{\rm i}\phi}$\\
   \hline
    $e^{{\rm i}\psi}$ & $e^{{\rm i}\phi}$ & $e^{{\rm i}\psi}$ & $e^{{\rm i}\phi}$\\
   \hline
    $e^{{\rm i}\psi}$ & $e^{{\rm i}\phi}$ & $e^{{\rm i}\phi}$ & $e^{{\rm i}\psi}$\\
   \hline
\end{tabular}
\end{center}
\caption{Solutions of (\ref{eq: Z_2 sols}) when all $r_j$'s are $\frac{1}{2}$.}
\label{tab:all4equal-Case1}
\end{table}
\noindent with $0\leq \phi,\psi< 2\pi$. When $\phi=0=\psi$, the simplest unitary operator is $$R_{ijk}=\frac{1}{2}\left(B_iB_jB_k+A_iA_jB_k+A_iB_jB_k+B_iA_jA_K\right).$$ The solutions when three or two of the $r_j$'s are equal to each other are shown in Table \ref{tab:3or2equal-Case1}.
\begin{table}[h!]
\begin{center}
\begin{tabular}{|c|c|c|c|}
    \hline
   $\alpha_0$  & $\alpha_1$ & $\alpha_2$ & $\alpha_3$ \\
   \hline
   $te^{{\rm i}\gamma}$ & $\sqrt{1-3t^2}~e^{{\rm i}\theta}$ & $te^{{\rm i}\gamma}$ & $te^{{\rm i}\gamma}$\\
   \hline
    $s$ & $\sqrt{1-s^2}~e^{-{\rm i}\frac{\pi}{2}}$ & $se^{{\rm i}\frac{\pi}{2}}$ & $\sqrt{1-s^2}~e^{{\rm i}\pi}$\\
   \hline
\end{tabular}
\end{center}
\caption{Solutions of (\ref{eq: Z_2 sols}) when three or two of the $r_j$'s are equal.}
\label{tab:3or2equal-Case1}
\end{table}
Another situation which simplifies the equations \eqref{eq: Z_2 sols} is obtained when two of the four $r_j$'s are set to 0. Setting $r_1=r_2=0,$ we obtain the solution :
\begin{equation}
    \alpha_0 = \cos\phi~e^{\mathrm{i}\theta}~;~\alpha_3 = \sin\phi~e^{\mathrm{i}\left(\theta+\frac{\pi}{2}\right)}.
\end{equation}
A more general solution to \eqref{eq: Z_2 sols}, when all four $r_j$'s are unequal, requires numerical study. For this study we will work in the qubit representation for the local Hilbert spaces. In the qubit representation the matrices $A$ and $B$ take the form 
\begin{eqnarray}
    A=a_1~X + a_2~Y + a_3~Z~;~B=b_1~X + b_2~Y + b_3~Z,
\end{eqnarray}
with $X$, $Y$ and $Z$ being the 2 by 2 Pauli matrices. For $A$ and $B$ to satisfy the conditions of case 1, the parameters need to satisfy 
\begin{eqnarray}
\sum\limits_{j=1}^3~a_j^2=\sum\limits_{j=1}^3~b_j^2=1~;~\sum\limits_{j=1}^3~a_jb_j=0.
\end{eqnarray}
Note that we have omitted a term proportional to the 2 by 2 identity operator, $\mathbb{1}$ for the expressions of $A$ and $B$. This is indeed required for the operators $A$ and $B$ to satisfy the conditions of case 1. We find that the simplest choices of $A$ and $B$ can be taken as 
\begin{equation}
    A=X~;~B=Z.
\end{equation}
Every other choice is equivalent to these two by a similarity transformation $Q$ and hence we work with this choice for simplicity. 
In this case one solution obtained numerically is as follows :
\begin{eqnarray}
    & \alpha_0 = 0.743304~e^{-\mathrm{i}3.48761}~;~\alpha_1 = 0.489214~e^{\mathrm{i}1.40343}, & \nonumber \\
    & \alpha_2 = 0.369857~e^{\mathrm{i}4.64383}~;~\alpha_3 = 0.267161~e^{-\mathrm{i}0.781393}.  
\end{eqnarray}

\paragraph{Case 2 : $A^2=\mathbb{1},~B^2=0$ -}Unlike the previous case, $B$ now cannot be realized by a single Hermitian Clifford generator. However, we still can express $B$ as some linear combination of them, as we will see shortly. We start with the choice given by (\ref{eq:tetraOP}). A simple computation shows that $R_{ijk}$ can never be unitary. As can be verified easily, $A_iA_jB_k,~ A_iB_jA_k,~B_iA_jA_K$ and $B_iB_jB_k$ commute among themselves and all of them are nilpotent\footnote{An operator $\hat{O}$, is nilpotent if it satisfies $\hat{O}^n=0$ for $n>1$. }. However, a sum of nilpotent, commuting matrices is again a nilpotent matrix, as seen {\it via} a simple computation. Consider two matrices ${\cal M},{\cal N}$, satisfying ${\cal M}^2=0={\cal N}^2,~ {\cal M}{\cal N}={\cal N}{\cal M}$. The matrix ${\cal P}={\cal M}+{\cal N}$ is nilpotent as
\begin{eqnarray}
{\cal P}^3&=&\left({\cal M}+{\cal N}\right)^2\left({\cal M}+{\cal N}\right)=2{\cal MN}\left({\cal M}+{\cal N}\right)=2{\cal NM}^2+2{\cal MN}^2=0.
\end{eqnarray}
This demonstrates that $R_{ijk}$, as given by (\ref{eq:tetraOP}), is also nilpotent and consequently cannot be made unitary. 

However this is not the case for the other tetrahedron operator (\ref{eq:tetraOPBA}), obtained by interchanging $A$ and $B$ in \eqref{eq:tetraOP}. The presence of the invertible term $A_iA_jA_k$ stops us from applying the previous lines of arguments.  First we choose $R_{ijk}$ as
\begin{eqnarray}\label{eq: case2(1,2_3,0)}
   R_{ijk}=a e^{{\rm i}\theta_a} B_i B_j A_k+b e^{{\rm i}\theta_b} B_iA_j B_k+c e^{{\rm i}\theta_c} A_i B_j B_k+d e^{{\rm i}\theta_d} A_iA_jA_k~;~a,b,c,d\in\mathbb{R}, \nonumber \\
\end{eqnarray}
where the complex coefficients are explicitly written in terms of real and imaginary parts. A simple way to check if the above $R_{ijk}$ is unitary or not is to analyze the eigenvalues of $R_{ijk}^{\dagger}R_{ijk}$. The necessary and sufficient condition for the unitarity is to have all the eigenvalues of $R_{ijk}^{\dagger}R_{ijk}$ as unity and this is true iff $a=b=c=0$ and $d=\pm1$. To show this we split $R_{ijk}$ into two operators 
\begin{eqnarray}
    {\cal J}_{ijk}=d e^{{\rm i}\theta_d} A_iA_jA_k,\quad {\cal K}_{ijk}=a e^{{\rm i}\theta_a} B_i B_j A_k+b e^{{\rm i}\theta_b} B_iA_j B_k+c e^{{\rm i}\theta_c} A_i B_j B_k.
\end{eqnarray}
They satisfy ${\cal JK}={\cal KJ}$, ${\cal J}^{\dagger}{\cal J}=d^2{\mathbb 1}$. Also, from the previous result, we identify ${\cal K}$ to be nilpotent. Then we have
\begin{eqnarray}
    R^{\dagger}R&=&\left({\cal J+K}\right)^{\dagger}\left({\cal J+K}\right),\nonumber\\
    &=&d^2{\mathbb 1}+\left({\cal J}^{\dagger}{\cal K}+{\cal K}^{\dagger}{\cal J}+{\cal K}^{\dagger}{\cal K}\right).
\end{eqnarray}

Clearly if we wish to have unitary $R$, we must have ${\cal L}=\left({\cal J}^{\dagger}{\cal K}+{\cal K}^{\dagger}{\cal J}+{\cal K}^{\dagger}{\cal K}\right)$ such that all its eigenvalues are identical, i.e. ${\cal L}$ should have only one distinct eigenvalue. On the other hand, we also have ${\cal L}^{\dagger}={\cal L}$, suggesting that ${\cal L}$ is a normal matrix as well, i.e. it satisfies $\left[{\cal L},{\cal L}^{\dagger}\right]=0$. Since a normal matrix is always unitarily diagonalizable and given that all its eigenvalues are equal, the matrix ${\cal L}$ has to be a constant multiple of the identity matrix.

Now if ${\cal L}= \eta {\mathbb 1}$, we should have ${\cal L}|\psi\rangle=\eta|\psi\rangle$ for arbitrary $|\psi\rangle$. Let us choose $|\psi\rangle$ to be a non-zero eigenvector of both ${\cal K}$ and ${\cal J}$, which is possible since $\left[{\cal J},{\cal K}\right]=0$. This obviously has zero eigenvalue under ${\cal K}$ as the latter is nilpotent. Then we have
\begin{eqnarray}
    {\cal L}|\psi\rangle={\cal K}^{\dagger}{\cal J}|\psi\rangle=\lambda_{\psi}{\cal K}^{\dagger}|\psi\rangle,\quad {\cal J}|\psi\rangle=\lambda_{\psi}|\psi\rangle.
\end{eqnarray}
Clearly this leads to the eigenvalue equation for ${\cal K}^{\dagger}$ as $${\cal K}^{\dagger}|\psi\rangle=\left(\frac{\eta}{\lambda_{\psi}}\right)|\psi\rangle.$$ However we know that ${\cal K}^{\dagger}$ is nilpotent and ${\cal J}^2\sim {\mathbb 1}$, indicating $\lambda_{\psi}\sim\pm 1$. Therefore the only conclusion is that $\eta=0$, i.e. ${\cal L}$ has to be the zero-matrix, leading to ${\rm Tr}\left[{\cal L}\right]=0$. Since $\left[{\cal J},{\cal K}\right]=0$, the expressions ${\cal J}^{\dagger}{\cal K},~{\cal K}^{\dagger}{\cal J}$ also are nilpotent and we arrive at
\begin{eqnarray}
    {\rm Tr}\left[{\cal L}\right]={\rm Tr}\left[{\cal K}^{\dagger}{\cal K}\right]=0~,
\end{eqnarray}
 which essentially tells that ${\cal K}$ also has to be the zero-matrix. This is easily seen by expanding the above expression as
\begin{eqnarray}
    {\rm Tr}\left[{\cal K}^{\dagger}{\cal K}\right]=\sum_i\left({\cal K}^{\dagger}{\cal K}\right)_{ii}=\sum_i\sum_j|{\cal K}_{ij}|^2~,
\end{eqnarray}
which equals to zero only if every element ${\cal K}_{ij}$ is zero. Clearly, this requires $a, b, c$ to all be exactly zero in the expression (\ref{eq: case2(1,2_3,0)}) for $R^\dag R\propto\mathbb{1}$. This proves the assertion that the tetrahedron operator in \eqref{eq:tetraOPBA} cannot be made unitary even though it is invertible.

An example of a qubit representation is given by the choice $A=Z,~B=X+{\rm i}Y$, satisfying $AB=-BA,~A^{-1}=A^{\dagger}=A,~B^2=0$. Same condition also holds if $B=X+{\rm i}Y$ is replaced with $B=X-{\rm i} Y$.

\paragraph{Case 3 : $A^2=A,~B^2=B$,~$AB=0=BA$ -} As already mentioned before, for this choice both (\ref{eq:tetraOP}) and (\ref{eq:tetraOPBA}) give non-invertible solutions. However, one crucial observation is that, by virtue of the relation $AB=0=BA$, both $A,~B$ anticommute as well as commute. This allows us to combine the identity operator along with the orthogonal projectors to form a commuting set on the local Hilbert space $V$. Therefore one can construct a total of $3^3=27$ different possible operators acting on $V\otimes V\otimes V$. All of them commute with each other and thus an arbitrary linear combination of them solves the tetrahedron equation. Thus the most general combination that one can construct is
\begin{eqnarray}\label{eq: projsol}
R_{ijk}&=&\sum_{m,m,\ell=1}^{3}\alpha_{m,n,\ell} (A_m)_i(A_n)_j(A_{\ell})_k~~,
    \end{eqnarray}
where $A_1=\mathbb{I},A_{2}=\Pi^{-},A_{3}=\Pi^{+}.$ A qubit realization for these orthogonal projectors are :
\begin{eqnarray}
    \Pi^+=\frac{{\mathbb 1}+Z}{2},\quad \Pi^-=\frac{{\mathbb 1}-Z}{2}.
\end{eqnarray}
Working in the basis of the Pauli $Z$ matrix, this operator is diagonal.  Any other pair of orthogonal projectors can be achieved by rotating $\Pi^{\pm}$ appropriately. We can construct unitary operators by tuning the coefficients $\alpha$'s accordingly. For example, the choice $\frac{1}{8}\exp\left({\rm i}\Phi\right);~j\in\{1,\cdots,27\},~0\leq\Phi<2\pi$ for all the coefficients, yields a unitary solution. 

\section{Tetrahedron operators from Yang-Baxter solutions} 
\label{sec:TetraFromYBsolutions}
The technique we will use to construct the tetrahedron operators are adapted from similar methods employed in \cite{Hietarinta_1993} where some tetrahedron solutions to the vertex form of the constant tetrahedron equations are considered. The method itself is rather simple and can be seen as a generalization of the Clifford solutions developed in \cite{padmanabhan2024solving}. In this section we will denote the tetrahedron operators by $T$ and the Yang-Baxter operators by $Y$\footnote{This is not to be confused with the second Pauli matrix, found in other sections of this paper.}, deviating from the notation used until now. Consider the following operators acting on $V\otimes V\otimes V$, 
\begin{eqnarray}
   & T = Y\otimes M~;~T_{ijk} = Y_{ij}M_k~, & \label{eq:tetraYM}\\
   & T = M\otimes Y~;~T_{ijk} = M_iY_{jk}. \label{eq:tetraMY}&
\end{eqnarray}
Here $M$ is an operator acting on the single space $V$. Substituting each of these ansatzse into the vertex form of the tetrahedron equation \eqref{eq:tetraVertexForm} reveals the conditions required on $Y$ and $M$ to make $T$ a solution. We find that when
\begin{eqnarray}
    Y_{12}Y_{13}Y_{23}=Y_{23}Y_{13}Y_{12}~;~[Y, M\otimes M]=0,
\end{eqnarray}
$T$ is a tetrahedron operator. These conditions can be generalized to
\begin{eqnarray}
    Y_{12}Y_{13}Y_{23}=\alpha~Y_{23}Y_{13}Y_{12}~;~M_1M_2Y_{12}= \frac{1}{\alpha}~Y_{12}M_1M_2.
\end{eqnarray}
For generic $\alpha\in\mathbb{C}$ the operator $Y$ satisfies a generalized version of the Yang-Baxter equation. The case when $\alpha=-1$, called the anti-Yang-Baxter equation, was first discussed in \cite{padmanabhan2024solving}. When $V=\mathbb{C}^2$, the qubit case, we do not find any non-trivial solution for a generic $\alpha$. Since this is the case considered in this paper we only consider the $\alpha=1$ case here. The generic $\alpha$ case will become relevant for higher dimensional representations ($V=\mathbb{C}^n$ for $n>2$). This technique of generating higher simplex operators by lifting lower simplex operators goes through for other cases as well. This is briefly discussed in Appendix \ref{app:lower2higher}.

In what follows we will consider two ways of constructing such tetrahedron operators. The first set is constructed out of Hietarinta's constant 4 by 4 Yang-Baxter operators \cite{Hietarinta_1992}. The second set of solutions uses the Yang-Baxter operators constructed out of Clifford algebras as outlined in \cite{padmanabhan2024solving}. 

\subsection{Tetrahedron operators from Hietarinta's Yang-Baxter solutions} 
\label{subsec:TetrafromHietarinta}
Consider the Yang-Baxter operator given by
\begin{equation}
    Y=\begin{pmatrix} Y_{00,\,00} & Y_{00,\,01} & Y_{00,\,10} & Y_{00,\,11} \\
Y_{01,\,00} & Y_{01,\,01} & Y_{01,\,10} & Y_{01,\,11} \\
Y_{10,\,00} & Y_{10,\,01} & Y_{10,\,10} & Y_{10,\,11} \\
Y_{11,\,00} & Y_{11,01} & Y_{11,\,10} & Y_{11,\,11} \end{pmatrix}.
\label{eq:matrixR}
\end{equation}
This satisfies the constant Yang-Baxter equation
\begin{equation}\label{eq:constantYBE}
    Y_{j_1\,j_2,\,k_1\,k_2}Y_{k_1\,j_3,\,l_1\,k_3}Y_{k_2\,k_3,\,l_2\,l_3}= Y_{j_2\,j_3,\,k_2\,k_3} Y_{j_1\,k_3,\,k_1\,l_3} Y_{k_1\,k_2,\,l_1\,l_2}.
\end{equation}
Hietarinta classified the constant 4 by 4 Yang-Baxter solutions \cite{Hietarinta_1992} up to continuous gauge transformations
\begin{equation}
    Y \to \kappa (Q\otimes Q)Y(Q\otimes Q)^{-1}\,,
\label{eq:Rchange}
\end{equation}
with $\kappa$ a complex factor and $Q$ an invertible $2\times 2$ matrix, and discrete transformations given by
\begin{eqnarray}
    & & Y_{i\,j,\,k\,l}\to Y_{k\,l,\,i\,j} \,,  \label{eq:discrete-1} \\
& & Y_{i\,j,\,k\,l} \to Y_{\bar{i}\,\bar{j},\,\bar{k}\,\bar{l}} \,, \label{eq:discrete-2} \\
& & Y_{i\,j,\,k\,l} \to Y_{j\,i,\,l\,k}. \label{eq:discrete-3}
\end{eqnarray}
Here \eqref{eq:discrete-1} means the matrix transpose (reflection about the main diagonal) taken in \eqref{eq:matrixR}, and $\bar{i}$ is the negation of $i$ (reflection about the second and third row and the second and third column), i.e., 
$\bar{0}\equiv 1$ and $\bar{1}\equiv 0$ in \eqref{eq:discrete-2}. Up to these equivalences the constant 4 by 4 Yang-Baxter operators fall into 10 classes :
\begin{eqnarray}
    & & Y_{H3,1} = \begin{pmatrix} k & 0 & 0 & 0 \\ 0 & 0 & p & 0 \\ 0 & q & 0 & 0 \\ 0& 0 & 0 & s \end{pmatrix} ,~
Y_{H2,1}=\begin{pmatrix} k^2 & 0 & 0 & 0 \\ 0 & k^2-pq & kp & 0 \\ 0 & kq & 0 & 0 \\ 0 & 0 & 0 & k^2 \end{pmatrix} , \nn\\
& & Y_{H2,2}=\begin{pmatrix} k^2 & 0 & 0 & 0 \\ 0 & k^2-pq & kp & 0 \\ 0 & kq & 0 & 0 \\ 0 & 0 & 0 & -pq \end{pmatrix} ,~
Y_{H2,3}=\begin{pmatrix} k & p & q & s \\ 0 & 0 & k & p \\ 0 & k & 0 & q\\ 0 & 0 & 0 & k \end{pmatrix}, \nn\\
& & Y_{H1,1}=\begin{pmatrix} p^2+2pq-q^2 & 0 & 0 & p^2-q^2 \\ 0 & p^2-q^2 & p^2+q^2 & 0 \\ 0 & p^2+q^2 & p^2 -q^2 & 0 \\ p^2-q^2 & 0 & 0 & p^2-2pq-q^2 \end{pmatrix}, \nn \\
& & Y_{H1,2}=\begin{pmatrix} p & 0 & 0 & k \\ 0 & p-q & p & 0 \\ 0 & q & 0 & 0 \\ 0 & 0 & 0 & -q \end{pmatrix}, ~
Y_{H1,3} =\begin{pmatrix} k^2 & -kp & kp & pq \\ 0 & 0 & k^2 & kq \\ 0 & k^2 & 0 & -kq \\ 0 & 0 & 0 & k^2 \end{pmatrix},~ 
Y_{H1,4}=\begin{pmatrix} 0 & 0 & 0 & p \\ 0 & k & 0 & 0 \\ 0 & 0 & k & 0 \\ q & 0 & 0 & 0 \end{pmatrix},
\nn \\
& & Y_{H0,1}=\begin{pmatrix} 1 & 0 & 0 & 1 \\ 0 & 0 & -1 & 0 \\ 0 & -1 & 0 & 0 \\ 0 & 0 & 0 & 1 \end{pmatrix}, ~
Y_{H0,2} = \begin{pmatrix} 1 & 0 & 0 & 1 \\ 0 & 1 & 1 & 0 \\ 0 & -1 & 1 & 0 \\ -1 & 0 & 0 & 1 \end{pmatrix}. 
\label{eq:Hietarinta4by4}
\end{eqnarray}
The 10 classes shown here comprise the invertible Yang-Baxter operators. They satisfy the braided form of the constant Yang-Baxter equation
\begin{equation}
    Y_{12}Y_{23}Y_{12} = Y_{23}Y_{12}Y_{23}.
\end{equation}
They are made to satisfy the vertex form of the constant Yang-Baxter equation in \eqref{eq:constantYBE} by the transformation
\begin{equation}
    Y \rightarrow PY,
\end{equation}
where $P$ is the permutation operator on $V\otimes V$. For the canonical basis of $V=\mathbb{C}^2$, $P$ takes the form
\begin{equation}\label{eq:permutationOP}
    P = \begin{pmatrix}
        1 & 0 & 0 & 0 \\
        0 & 0 & 1 & 0 \\
        0 & 1 & 0 & 0 \\
        0 & 0 & 0 & 1
    \end{pmatrix}.
\end{equation}
Apart from these 10 classes we also include the permutation operator \eqref{eq:permutationOP} as part of the constant invertible 4 by 4 Yang-Baxter operators. The full classification of 4 by 4 constant solutions also include non-invertible Yang-Baxter operators \cite{Hietarinta_1992}. They can still be used to construct non-invertible tetrahedron operators using \eqref{eq:tetraYM} and \eqref{eq:tetraMY}. It should be noted that non-invertible constant Yang-Baxter and tetrahedron can still be helpful to obtain invertible Yang-Baxter and tetrahedron operators that depend on spectral parameters. These play a significant role in obtaining integrable models.

The tetrahedron operators corresponding to these 10 classes are obtained after identifying the 2 by 2 $M$ matrix that satisfies $[Y, M\otimes M]=0$. 
Let $M=\begin{pmatrix} m_1 & m_2  \\ m_3 & m_4  \end{pmatrix}$ with $m_1,m_2,m_3,m_4\in\mathbb{C}$. We require $M$ to be invertible to obtain invertible tetrahedron operators. Then the solutions for all classes under enforcement of the condition on the parameters of $M$, excluding $Y$ are the following :
\paragraph{(1) $Y_{H3,1}$: } In this case
\begin{equation}\label{eq:M31}
    M=\begin{pmatrix} m_1 & 0  \\ 0 & m_4  \end{pmatrix}.
\end{equation}
This is invertible as long as $m_1,m_4\neq 0$.

\paragraph{(2) $Y_{H2,1}$: } We have three possible solutions in this case :
\begin{equation}\label{eq:M21}
    M=\left\{\begin{pmatrix} m_1 & 0  \\ 0 & m_4  \end{pmatrix},~~\begin{pmatrix} 0 & 0  \\ m_3 & 0  \end{pmatrix},~~\begin{pmatrix} 0 & m_2  \\ 0 & 0  \end{pmatrix}\right\}.
\end{equation}
We use just the first of these three matrices as the other two are nilpotent and hence non-invertible. The first is invertible when $m_1,m_4\neq 0$.

\paragraph{(3) $Y_{H2,2}$: } The choice for $M$ is the same as the previous two cases. Invertible matrices when $m_1,m_4\neq 0$ :
\begin{equation}\label{eq:M22}
    M=\begin{pmatrix} m_1 & 0  \\ 0 & m_4  \end{pmatrix}.
\end{equation}

\paragraph{(4) $Y_{H2,3}$: } We have two possibilities for invertible $M$ in this case :
\begin{equation}
    M=\left\{\begin{pmatrix} m_1 & m_2  \\ 0 & m_4  \end{pmatrix}, \begin{pmatrix} 1 & - \frac{2  s}{p+q}  \\ 0 & 1  \end{pmatrix}\right\}.
\end{equation}
However the second is a special case of the first and so we ignore it. This is invertible when $m_1\neq 0$.

\paragraph{(5) $Y_{H1,1}$: } The invertible $M$ matrix in this case is the third Pauli matrix,
\begin{equation}
    M=\begin{pmatrix} 1 & 0  \\ 0 & -1  \end{pmatrix} = Z.
\end{equation}
The other choices comprise of projectors and are hence non-invertible :
\begin{equation}
    M=\left\{\begin{pmatrix} 1 & \pm \frac{\sqrt{p-q}}{\sqrt{p+q}}  \\ \pm \frac{\sqrt{p-q}}{\sqrt{p+q}} & \frac{(p-q)}{p+q}\end{pmatrix},~~
\begin{pmatrix} 1 & \pm \frac{\sqrt{p-q}}{\sqrt{p+q}}  \\ \mp \frac{\sqrt{p-q}}{\sqrt{p+q}} & -\frac{(p-q)}{p+q}\end{pmatrix}\right\},~~ M^2=M.
\end{equation}

\paragraph{(6) $Y_{H1,2}$: } As in the $H(1,1)$ class, the invertible $M$ is just the third Pauli matrix, $M=Z$. The remaining two options are projectors
$$M=\begin{pmatrix} \pm \frac{\sqrt{p+q}}{\sqrt{k}} & 1 \\ 0 & 0  \end{pmatrix}.$$

\paragraph{(7) $Y_{H1,3}$: } For this class we obtain one invertible $M$ matrix that is similar to one obtained in the $H(2,3)$ class. The other possibility is non-invertible.
\begin{equation}
    M=\left\{\begin{pmatrix} m_1 & m_2  \\ 0 &  m_1  \end{pmatrix}, \begin{pmatrix} 0& m_2 \\ 0 & 0  \end{pmatrix}\right\}.
\end{equation}

\paragraph{(8) $Y_{H1,4}$: } There are two possible non-trivial invertible $M$ matrices in this case. One of them is just the third Pauli matrix, $Z$, while the other is a linear combination of the first two Pauli matrices $X$ and $Y$ :
\begin{equation}\label{eq:M14}
    M=\left\{\begin{pmatrix} 1 & 0  \\ 0 & -1  \end{pmatrix}, \begin{pmatrix} 0& 1 \\ \pm \frac{\sqrt{q}}{\sqrt{p}} & 0  \end{pmatrix}\right\},~~M^2 \propto \mathbb{I}.
\end{equation}

\paragraph{(9) $Y_{H0,1}$: } The only invertible choice for $M$ is the third Pauli matrix $Z$. The other choice is a nilpotent $M$ that is non-invertible.

\paragraph{(10) $Y_{H0,2}$: } This is similar to the $H(0,1)$ class, with $M=Z$ being the only invertible option.

\paragraph{(11) $P_{12}$: } This presents the largest class of solutions as there is no restriction on the $M$ matrices. This follows from the defining property of the permutation matrices. To obtain invertible $M$ matrices we choose $M\in GL(2,\mathbb{C})$. 

\paragraph{Unitary solutions -} We will now identify the unitary tetrahedron operators among the 11 classes of solutions derived from Hietarinta's Yang-Baxter operators. As each tetrahedron operator is of the form $Y\otimes M$ or $M\otimes Y$ we require both $Y$ and $M$ to be unitary separately. We are aided by the 5 families of unitary 4 by 4 constant Yang-Baxter operators deduced in \cite{dye2003unitarysolutionsyangbaxterequation}. We will use the results from this paper for making the $Y$ part unitary. In each case we will see the modifications in the $Q$ or $M$ matrices inherited from making the single qubit $M$ matrices unitary. As discussed in Sec. \ref{sec:preliminaries} we specify the unitary families, $$ \kappa~\mathcal{Q}~\left(Y\otimes M\right)~\mathcal{Q}^{-1}~~;~~\kappa~\mathcal{Q}~\left(M\otimes Y\right)~\mathcal{Q}^{-1}~,$$ by determining the $Q$ matrices. Recall that $\mathcal{Q}=Q\otimes Q\otimes Q$. The Hietarinta families $H1,1$ and $H0,1$ produce no unitary solutions even after conjugation \cite{dye2003unitarysolutionsyangbaxterequation} and hence we omit these cases. The remaining 9 classes, including the permutation operator, are discussed next. 
\begin{enumerate}
    \item {\it {\bf Family 1 :}} The Yang-Baxter operator corresponds to the $H3,1$ Hietarinta class \eqref{eq:Hietarinta4by4}. This operator and the corresponding $Q$ matrix are given by
    \begin{equation}
        Y = \begin{pmatrix}
            1 & 0 & 0 & 0 \\
            0 & p & 0 & 0 \\
            0 & 0 & q & 0 \\
            0 & 0 & 0 & r
        \end{pmatrix},~~Q = \begin{pmatrix}
            q_1 & q_2 \\ -\frac{q_1\bar{q_2}}{\bar{q_4}} & q_4
        \end{pmatrix}.
    \end{equation}
    The parameters of the $Y$ matrix are phases, $ |p|=|q|=|r|=1.$ This makes $Y$ a unitary matrix by itself and also ensures that the family $\left(Q\otimes Q\right)~Y~\left(Q^{-1}\otimes Q^{-1}\right)$ stays unitary as $$ \left(Q^\dag Q\otimes Q^\dag Q\right)Y^{-1} =Y^\dag\left(Q^\dag Q\otimes Q^\dag Q\right).$$
    This follows from the fact that $Q^\dag Q$ is diagonal. For the Hietarinta class $H3,1$, the $M$ operator is diagonal \eqref{eq:M31}. To identify the unitary family of such $M$ matrices we require 
    $$ Q^\dag QM^{-1} = M^\dag Q^\dag Q.$$ The diagonal nature of $Q^\dag Q$ at $q_3=-\frac{q_1\bar{q_2}}{q_4}$ reduces $M$ to just the Pauli $Z$ operator. Thus the two sets of unitary tetrahedron operators we obtain from Hietarinta's $H3,1$ class are just 
    \begin{eqnarray}\label{eq:unitaryHFamily-1}
        \mathcal{Q}\begin{pmatrix}
            1 & 0 & 0 & 0 \\
            0 & p & 0 & 0 \\
            0 & 0 & q & 0 \\
            0 & 0 & 0 & r
        \end{pmatrix}\otimes \begin{pmatrix}
            1 & 0 \\ 0 & -1
        \end{pmatrix}\mathcal{Q}^{-1}~~;~~\mathcal{Q}\begin{pmatrix}
            1 & 0 \\ 0 & -1
        \end{pmatrix}\otimes  \begin{pmatrix}
            1 & 0 & 0 & 0 \\
            0 & p & 0 & 0 \\
            0 & 0 & q & 0 \\
            0 & 0 & 0 & r
        \end{pmatrix}\mathcal{Q}^{-1},
    \end{eqnarray}
with $p$, $q$ and $r$ being mere phases and $Q = \begin{pmatrix}
            q_1 & q_2 \\ -\frac{q_1\bar{q_2}}{\bar{q_4}} & q_4
        \end{pmatrix}$. The Yang-Baxter operators corresponding to the Hietarinta classes $H2,1$, $H2,2$ and $H1,2$ are not unitary by themselves. Under conjugation they produce unitary solutions that are already captured by the $H3,1$ class (Family 1). A similar scenario occurs for the $H2,3$ and $H1,3$ Hietarinta classes, which under conjugation, reduce to the trivial identity operator that is unitary. Thus these two classes also fall under Family 1 generated by the $H3,1$ Hietarinta class. 
    \item {\it {\bf Family 2 :}} The Yang-Baxter operator corresponding to the $H0,2$ \eqref{eq:Hietarinta4by4} Hietarinta class becomes unitary when scaled by the constant $\frac{1}{\sqrt{2}}$. The corresponding Yang-Baxter operator and the $Q$ matrix is given by
    \begin{equation}
        \tilde{Y}_{H0,2} = \begin{pmatrix} \frac{1}{\sqrt{2}} & 0 & 0 & \frac{1}{\sqrt{2}} \\ 0 & \frac{1}{\sqrt{2}} & \frac{1}{\sqrt{2}} & 0 \\ 0 & \frac{1}{\sqrt{2}} & -\frac{1}{\sqrt{2}} & 0 \\ -\frac{1}{\sqrt{2}} & 0 & 0 & \frac{1}{\sqrt{2}} \end{pmatrix},~~Q = \begin{pmatrix}
            q_1 & q_2 \\ -\frac{q_1\bar{q_2}}{\bar{q_4}} & q_4
        \end{pmatrix}, |q_1|=|q_4|.
    \end{equation}
    In this case $M=Z$, the third Pauli matrix. As $Q^\dag Q$ is diagonal at $q_3=-\frac{q_1\bar{q_2}}{\bar{q_4}}$, it satisfies the condition $Q^\dag Q M^{-1}=M^\dag Q^\dag Q$. Thus the unitary families of tetrahedron operators are given by 
    \begin{eqnarray}\label{eq:unitaryHFamily-2}
        \mathcal{Q}\left[\tilde{Y}_{H0,2}\otimes Z\right]\mathcal{Q}^{-1}~;~  \mathcal{Q}\left[Z\otimes\tilde{Y}_{H0,2}\right]\mathcal{Q}^{-1},
    \end{eqnarray}
    with $Q = \begin{pmatrix}
            q_1 & q_2 \\ -\frac{q_1\bar{q_2}}{\bar{q_4}} & q_4
        \end{pmatrix}$ and $|q_1|=|q_4|$.
    \item {\it {\bf Family 3 :}} This is generated by the Yang-Baxter operator in the $H1,4$ Hietarinta class \eqref{eq:Hietarinta4by4}. The unitary family obtained from this Yang-Baxter operator and the associated $Q$ matrices is given by
    \begin{eqnarray}
        Y = \begin{pmatrix}
            0 & 0 & 0 & p \\
            0 & 0 & 1 & 0 \\
            0 & 1 & 0 & 0 \\
            q & 0 & 0 & 0
        \end{pmatrix},~~Q=\begin{pmatrix}
            q_1 & q_2 \\ -\frac{q_1\bar{q_2}}{\bar{q_4}} & q_4
        \end{pmatrix},
    \end{eqnarray}
    with the conditions that $|q_1|^2=|q||q_4|^2$ and $|q_4|^2=|p||q_1|^2$. There are two possible choices for the $M$ matrices in this case \eqref{eq:M14}. When $M=Z$, there are no further restrictions on $Y$ and $Q$ as $Q^\dag Q$ is diagonal and it commutes with $Z$. Thus we obtain two types of unitary families of tetrahedron operators 
    \begin{eqnarray}\label{eq:unitaryHFamily-3-1}
        \mathcal{Q}\left[\begin{pmatrix}
            0 & 0 & 0 & p \\
            0 & 0 & 1 & 0 \\
            0 & 1 & 0 & 0 \\
            q & 0 & 0 & 0
        \end{pmatrix}\otimes Z \right]\mathcal{Q}^{-1}~;~ \mathcal{Q}\left[Z\otimes \begin{pmatrix}
            0 & 0 & 0 & p \\
            0 & 0 & 1 & 0 \\
            0 & 1 & 0 & 0 \\
            q & 0 & 0 & 0
        \end{pmatrix} \right]\mathcal{Q}^{-1},
    \end{eqnarray}
    with $|q_1|^2=|q||q_4|^2$ and $|q_4|^2=|p||q_1|^2$. When $M=\begin{pmatrix}
        0 & 1 \\ \pm \frac{\sqrt{q}}{\sqrt{p}} & 0 
    \end{pmatrix}$, we find that the condition $Q^\dag QM^{-1} = M^\dag Q^\dag Q$ makes $|q_1|=|q_4|$, restricting both $p$ and $q$ to phases. Thus the tetrahedron operators in this case become 
    \begin{eqnarray}\label{eq:unitaryHFamily-3-2}
        & \mathcal{Q}\left[\begin{pmatrix}
            0 & 0 & 0 & e^{\mathrm{i}\theta_p} \\
            0 & 0 & 1 & 0 \\
            0 & 1 & 0 & 0 \\
            e^{\mathrm{i}\theta_q} & 0 & 0 & 0
        \end{pmatrix}\otimes \begin{pmatrix}
        0 & 1 \\ \pm \frac{\sqrt{e^{\mathrm{i}\theta_q}}}{\sqrt{e^{\mathrm{i}\theta_p}}} & 0 
    \end{pmatrix}\right] \mathcal{Q}^{-1}~, & \nonumber \\
    & \mathcal{Q}\left[\begin{pmatrix}
        0 & 1 \\ \pm \frac{\sqrt{e^{\mathrm{i}\theta_q}}}{\sqrt{e^{\mathrm{i}\theta_p}}} & 0 
    \end{pmatrix}\otimes\begin{pmatrix}
            0 & 0 & 0 & e^{\mathrm{i}\theta_p} \\
            0 & 0 & 1 & 0 \\
            0 & 1 & 0 & 0 \\
            e^{\mathrm{i}\theta_q} & 0 & 0 & 0
        \end{pmatrix}\right] \mathcal{Q}^{-1}~, &
        \end{eqnarray} with $Q = \begin{pmatrix}
            |q_1|e^{\mathrm{i}\theta_1} & q_2 \\ -\bar{q_2}e^{\mathrm{i}(\theta_1+\theta_4)} & |q_1|e^{\mathrm{i}\theta_4}
        \end{pmatrix}$.
    \item {\it {\bf Family 4 :}} This family is also generated by the $H1,4$ Hietarinta class \eqref{eq:Hietarinta4by4}. In this case there is no restriction on the invertible $Q$ matrix. The parameters of the Yang-Baxter operator are related to the parameters of the $Q$ matrix as
    \begin{eqnarray}
       & p = \frac{(\bar{q_2}q_2+\bar{q_4}q_4)(\bar{q_1}q_2+\bar{q_3}q_4)}{(\bar{q_1}q_1+\bar{q_3}q_3)(\bar{q_2}q_1+\bar{q_4}q_3)} = \frac{y\bar{z}}{xz}~, & \nonumber \\
        & q = \frac{(\bar{q_1}q_1+\bar{q_3}q_3)(\bar{q_2}q_1+\bar{q_4}q_3)}{(\bar{q_2}q_2+\bar{q_4}q_4)(\bar{q_1}q_2+\bar{q_3}q_4)} = \frac{xz}{y\bar{z}} =\frac{1}{p}. &
    \end{eqnarray}
    The freedom in $Q$ collapses when $M=Z$ as we see that it has to be diagonal ($q_3=-\frac{q_1\bar{q_2}}{\bar{q_4}}$) for $Q^\dag Q$ to commute with $Z$. Furthermore this condition implies that $z=0$, making both $p$ and $q$ ill-defined. Thus we cannot obtain unitary tetrahedron operator by appending $M=Z$ to the above Yang-Baxter operator. However there is a way around this as shown in \eqref{eq:unitaryHFamily-3-1}. This is precisely part of the third family. On the other hand when $M=\begin{pmatrix}
        0 & 1 \\ \pm \frac{\sqrt{q}}{\sqrt{p}} & 0 
    \end{pmatrix}$, the condition $Q^\dag QM^{-1}=M^\dag Q^\dag Q$ makes $x=y$ and $z=\bar{z}$, that is it places the following relations on the parameters of $Q$:
    \begin{eqnarray}
        \bar{q_1}q_1 + \bar{q_3}q_3 & = & \bar{q_2}q_2 + \bar{q_4}q_4~, \nonumber \\
        \bar{q_2}q_1 \mp  \bar{q_1}q_2 & = & \bar{q_3}q_4 \mp \bar{q_4}q_3.
    \end{eqnarray}
As a result $p=q=1$ and we find the two sets of unitary tetrahedron operators as
    \begin{eqnarray}\label{eq:unitaryHFamily-4}
        & \mathcal{Q}\left[\begin{pmatrix}
            0 & 0 & 0 & 1 \\
            0 & 0 & 1 & 0 \\
            0 & 1 & 0 & 0 \\
            1 & 0 & 0 & 0
        \end{pmatrix}\otimes \begin{pmatrix}
        0 & 1 \\  \pm 1 & 0 
    \end{pmatrix}\right] \mathcal{Q}^{-1} &, \nonumber \\
    & \mathcal{Q}\left[\begin{pmatrix}
        0 &  1 \\ \pm 1 & 0 
    \end{pmatrix}\otimes \begin{pmatrix}
            0 & 0 & 0 & 1 \\
            0 & 0 & 1 & 0 \\
            0 & 1 & 0 & 0 \\
            1 & 0 & 0 & 0
        \end{pmatrix}\right] \mathcal{Q}^{-1}. &
    \end{eqnarray}
    \item {\it {\bf Family 5 :}} The fifth and largest family of unitary tetrahedron operators is generated with the permutation operator as the Yang-Baxter operator. In this case there is no further restriction on the $Q$ matrices apart from their invertibility. This follows from the defining property of the permutation operator, $[Q\otimes Q, P]=0$. Thus $Q\in GL(2,\mathbb{C})$. For the same reason there is no restriction on the $M$ matrices as well. Thus the unitary family is obtained with $M$ being an arbitrary element of U(2)  $$M=\begin{pmatrix}
    e^{-\frac{\mathrm{i}}{2}(\beta+\delta)}\cos{\frac{\gamma}{2}} & -e^{-\frac{\mathrm{i}}{2}(\beta-\delta)}\sin{\frac{\gamma}{2}} \\ 
    e^{\frac{\mathrm{i}}{2}(\beta-\delta)}\sin{\frac{\gamma}{2}} & e^{\frac{\mathrm{i}}{2}(\beta+\delta)}\cos{\frac{\gamma}{2}}
\end{pmatrix}.$$ This requirement however restricts the choice of $Q$ to $U(2)\subset GL(2,\mathbb{C})$. This trivially satisfies the condition $Q^\dag QM=MQ^\dag Q$ as $Q^\dag Q=\mathbb{1}$ when $Q\in U(2)$. Thus we find the unitary tetrahedron operators to be 
\begin{eqnarray}\label{eq:unitaryHFamily-5}
    \mathcal{Q}\left[P\otimes M\right]\mathcal{Q}^{-1}~;~\mathcal{Q}\left[M\otimes P\right]\mathcal{Q}^{-1},
\end{eqnarray}
with $M,Q\in U(2)$.
\end{enumerate}

\subsection{Tetrahedron operators from Clifford Yang-Baxter solutions} 
\label{subsec:TetrafromClifford}
We will now use Yang-Baxter operators constructed using Clifford algebras \cite{padmanabhan2024solving} to construct tetrahedron operators. We recall the procedure developed in \cite{padmanabhan2024solving}. Consider two anticommuting operators $A$ and $B$. Then the operator 
\begin{equation}
    \alpha~A_iA_j + \beta~B_iB_j, 
\end{equation}
satisfies the constant Yang-Baxter (2-simplex) equation for arbitrary complex parameters $\alpha$ and $\beta$. We will now tensor the above Yang-Baxter solution with an operator $C$, that plays the role of $M$ used in Sec.\ref{subsec:TetrafromHietarinta}. Then we require $C\otimes C$ to commute with the Yang-Baxter solution $\alpha~A_iA_j + \beta~B_iB_j$. This is achieved in one of two ways :
\begin{enumerate}
    \item when $\{C,A\}=\{C,B\}=0$.
    \item when $[C,A]=[C,B]=0$.
\end{enumerate}
Then the tetrahedron operators are given by
\begin{eqnarray}
    R_{ijk} & = &  \alpha~A_iA_jC_k + \beta~B_iB_jC_k~;~\alpha,\beta\in\mathbb{C}, \label{eq:tetraABC-1}\\
    R_{ijk} & = &  \alpha~C_iA_jA_k + \beta~C_iB_jB_k~;~\alpha,\beta\in\mathbb{C}. \label{eq:tetraABC-2}
\end{eqnarray}
These operators satisfy the vertex form of the tetrahedron equation \eqref{eq:tetraVertexForm}. The techniques presented in \cite{padmanabhan2024solving} can also be used to verify this claim. Other words in $A$, $B$ and $C$ satisfy generalizations of the tetrahedron equation (See Appendix \ref{app:3from2} for details). 

\paragraph{$\mathbf{\{A,B\}=\{C,A\}=\{C,B\}=0}$ :} To discuss the unitary solutions among the operators in \eqref{eq:tetraABC-1} and \eqref{eq:tetraABC-2} we consider the situations when three two by two matrices ($A$, $B$ and $C$) mutually anticommute with one another. This is possible in two situations :
\begin{enumerate}
    \item $A^2=B^2=C^2=\mathbb{1}$. These operators can be obtained from Clifford algebras of orders $\geq$ 3. It is also possible to have some of these operators square to $-\mathbb{1}$, but in such cases we scale them by $\mathrm{i}=\sqrt{-1}$.
    \item One of the three operators squares to $\mathbb{1}$ and the other two square to 0. Furthermore the other two also commute with each other as they are proportional to each other. There are three ways to achieve the stated relations :
    \begin{eqnarray}
        & A^2=\mathbb{1}~;~B^2=C^2=0,~BC =CB = 0, & \nonumber \\
        & B^2=\mathbb{1}~;~C^2=A^2=0,~CA =AC= 0, & \nonumber \\
        & C^2=\mathbb{1}~;~A^2=B^2=0,~AB =BA = 0. & 
    \end{eqnarray}
    It is enough to consider one of these sets as they can be rotated into one another by a similarity transformation. For example if we take the first situation, then we have $A^2=\mathbb{1}$ and $B$ and $C$ square to zero and are proportional to one another. This is precisely the same as Case 2 of the Clifford solutions considered in Sec. \ref{sec:unitaryCliffordSolutions}. There are no unitary solutions in this and so we do not consider this case further.
\end{enumerate}

\paragraph{Case 1 : $A^2=B^2=C^2=\mathbb{1}$.}
The operators $A$, $B$ and $C$ are taken to be generators of a Clifford algebra and hence are Hermitian. Then the inverse operators of \eqref{eq:tetraABC-1} and \eqref{eq:tetraABC-2} satisfy 
\begin{eqnarray}
    R^{-1}_{ijk}=R^\dag_{ijk} & = & \Bar{\alpha}~A_iA_jC_k + \Bar{\beta}~B_iB_jC_k ,\label{eq:tetraABC-1-inverse}\\
    R^{-1}_{ijk}=R^\dag_{ijk} & = & \Bar{\alpha}~C_iA_jA_k + \Bar{\beta}~C_iB_jB_k. \label{eq:tetraABC-2-inverse}
\end{eqnarray}
For unitary solutions the parameters satisfy
\begin{equation}
    |\alpha|^2+|\beta|^2=1~;~\alpha\Bar{\beta} + \Bar{\alpha}\beta = 0.
\end{equation}
Assuming $\alpha=|\alpha|e^{\mathrm{i}\theta_\alpha}$ and $\beta=|\beta|e^{\mathrm{i}\theta_\beta}$, we have $$ |\alpha|=\sin{\phi}~;~|\beta|=\cos{\phi},$$ and the condition on the arguments become
\begin{equation}
   \cos{(\theta_\alpha-\theta_\beta)}=0~\implies~\theta_\alpha = \theta_\beta + \frac{n\pi}{2}~;~n\in\mathbb{Z}.
\end{equation}
These representation independent solutions also represent the entire family of unitary operators, $\mathcal{Q}R\mathcal{Q}^{-1}$, as Remark 3.1 holds for this case well. This is seen as a consequence of the structure of the tetrahedron operators.

In the qubit representation the operators $A$, $B$ and $C$ take the form
\begin{eqnarray}
    A & = & a_1~X + a_2~Y + a_3~Z, \nonumber\\
    B & = & b_1~X + b_2~Y + b_3~Z, \nonumber\\
    C & = & c_1~X + c_2~Y + c_3~Z.
\end{eqnarray}
The parameters satisfy :
\begin{eqnarray}
&\sum\limits_{j=1}^3~a^2_j=\sum\limits_{j=1}^3~b^2_j=\sum\limits_{j=1}^3~c^2_j = 1, & \nonumber \\
&\sum\limits_{j=1}^3~a_jb_j=\sum\limits_{j=1}^3~b_jc_j=\sum\limits_{j=1}^3~c_ja_j=0. &
\end{eqnarray}
The simplest choice for the three operators are given by the three Pauli matrices 
\begin{equation}
    A=X,~B=Z,~C=Y.
\end{equation}
Every other solution is a rotation of this basis choice of the Pauli matrices. The tetrahedron operator in \eqref{eq:tetraABC-1} then takes the form
\begin{equation}
    R=
\begin{pmatrix}
 0 & -i \beta  & 0 & 0 & 0 & 0 & 0 & -i \alpha  \\
 i \beta  & 0 & 0 & 0 & 0 & 0 & i \alpha  & 0 \\
 0 & 0 & 0 & i \beta  & 0 & -i \alpha  & 0 & 0 \\
 0 & 0 & -i \beta  & 0 & i \alpha  & 0 & 0 & 0 \\
 0 & 0 & 0 & -i \alpha  & 0 & i \beta  & 0 & 0 \\
 0 & 0 & i \alpha  & 0 & -i \beta  & 0 & 0 & 0 \\
 0 & -i \alpha  & 0 & 0 & 0 & 0 & 0 & -i \beta  \\
 i \alpha  & 0 & 0 & 0 & 0 & 0 & i \beta  & 0 \\
\end{pmatrix}.
\end{equation}
This is unitary for  $$\alpha=\sin{(\phi)} e^{(i \theta)},~\beta=\cos{(\phi)} e^{\left(i (\frac{\pi}{2}+\theta)\right)}.$$

\paragraph{$\mathbf{\{A,B\}=[C,A]=[C,B]=0}$ :} In the case of 2 by 2 matrices, the only possible choice for $C$, that is invertible, is that it is proportional to the identity matrix when both $A$ and $B$ are also invertible. Thus this choice suggests that the resulting  tetrahedron operator ($R_{ijk}$) is just the Yang-Baxter operator acting on either the first two qubits ($i$ and $j$) or the last two qubits ($j$ and $k$). Their unitary versions do not give us any new unitary tetrahedron operator and so we will not consider this choice. 

When $A$ and $B$ are allowed to be non-invertible, that is they are orthogonal projectors (See Case 3 in Sec. \ref{sec:unitaryCliffordSolutions}), then we have a non-trivial choice for $C$ :
\begin{equation}
    A = \frac{\mathbb{1}+Z}{2}\equiv\Pi^+,~B = \frac{\mathbb{1}-Z}{2}\equiv\Pi^-,~C=\begin{pmatrix}
        a & 0 \\ 0 & d
    \end{pmatrix}.
\end{equation}
 Then the tetrahedron operator is given by
 \begin{eqnarray}\label{eq:tetrafromcase3YB}
     R_{ijk} & = & \left[ \alpha_1~\mathbb{1} + \alpha_2~\Pi^+_i + \alpha_3~\Pi^+_j +  \alpha_4~\Pi^-_i + \alpha_5~\Pi^-_j\right.\nonumber \\
     & + & \left. \alpha_6~\Pi^+_i\Pi^+_j + \alpha_7~\Pi^-_i\Pi^-_j 
      +  \alpha_8~\Pi^+_i\Pi^-_j + \alpha_9~\Pi^-_i\Pi^+_j \right]C_k.
\end{eqnarray}
Note that the term in the parantheses is a Yang-Baxter solution. The operator $C$ is appended after this, on the third qubit. The other tetrahedron operator is obtained by appending $C$ before this Yang-Baxter solution or on the first qubit. Note that $C$ is a linear combination of the 2 by 2 identity operator $\mathbb{1}$ and the third Pauli operator $Z$. This makes the solution the same as Case 3 of the Clifford tetrahedron solutions \eqref{eq: projsol}. For the special case of $C=Z$, the unitary version of the tetrahedron operator in \eqref{eq:tetrafromcase3YB} is given by 
\begin{equation}
R=
\begin{pmatrix}
 \beta _1 & 0 & 0 & 0 & 0 & 0 & 0 & 0 \\
 0 & -\beta _1 & 0 & 0 & 0 & 0 & 0 & 0 \\
 0 & 0 & \beta _2 & 0 & 0 & 0 & 0 & 0 \\
 0 & 0 & 0 & -\beta _2 & 0 & 0 & 0 & 0 \\
 0 & 0 & 0 & 0 & \beta _3 & 0 & 0 & 0 \\
 0 & 0 & 0 & 0 & 0 & -\beta _3 & 0 & 0 \\
 0 & 0 & 0 & 0 & 0 & 0 & \beta _4 & 0 \\
 0 & 0 & 0 & 0 & 0 & 0 & 0 & -\beta _4 \\
\end{pmatrix},
\end{equation}
with $\beta _1=\alpha _0+\alpha _1+\alpha _2+\alpha _8,~\beta _2=\alpha _0+\alpha _1+\alpha _4+\alpha _5,~\beta _3=\alpha _0+\alpha _2+\alpha _3+\alpha _6,~\beta _4=\alpha _0+\alpha _3+\alpha _4+\alpha _7,$ being phases. 

\begin{remark}
    The constant 4 by 4 invertible Yang-Baxter solutions are classified by the 11 classes, including the permutation operator, by Hietarinta \cite{Hietarinta_1992}. Given this, it is imperative to check if the tetrahedron operators in \eqref{eq:tetraABC-1} and \eqref{eq:tetraABC-2} are equivalent to the tetrahedron operators obtained from Hietarinta's Yang-Baxter solutions Sec. \ref{subsec:TetrafromHietarinta}. We discuss this for the two cases considered here :
    \begin{enumerate}
        \item {\bf $\mathbf{\{A,B\}=\{C,A\}=\{C,B\}=0}$ -} The Yang-Baxter operator $$ \alpha~X\otimes X + \beta~Z\otimes Z$$ reduces to a Yang-Baxter operator in the $H1,4$ Hietarinta class $$ Y_{H1,4} = \begin{pmatrix}
            0 & 0 & 0 & p \\ 0 & 0 & \alpha+\beta & 0 \\ 0 & \alpha+\beta & 0 & 0 \\ \frac{(\alpha-\beta)^2}{p} & 0 & 0 & 0
        \end{pmatrix}~,$$ under the conjugation by the $Q$ matrix given by
        \begin{equation}
            Q =\frac{1}{2} \left(
\begin{array}{cc}
 \frac{\sqrt{p}}{\sqrt{\alpha -\beta }} & \frac{i \sqrt{p}}{\sqrt{\alpha -\beta }} \\
 i & 1 \\
\end{array}
\right).
        \end{equation}
        There are two tetrahedron operators corresponding to the two different $M$ matrices for the $H1,4$ Hietarinta class \eqref{eq:M14}. To check if these two tetrahedron operators are equivalent to ones obtained from \eqref{eq:tetraABC-1} and \eqref{eq:tetraABC-2} we need to check the transformation of the $Y$ matrix under the conjugation of the above $Q$ matrix. We find that 
        \begin{eqnarray}
             Y \rightarrow QYQ^{-1} = -Z.
        \end{eqnarray}
        This is one of the $M$ matrices for the $H1,4$ Hietarinta class \eqref{eq:M14}. Thus we conclude that this tetrahedron operator falls in Family-3 \eqref{eq:unitaryHFamily-3-1}. 
        

        
        \item {\bf $\mathbf{\{A,B\}=[C,A]=[C,B]=0}$ -} In this case the Yang-Baxter operator  
        \begin{eqnarray}\label{eq:YijRemark4.1}
         Y_{ij} & = & \left[ \alpha_1~\mathbb{1} + \alpha_2~\Pi^+_i + \alpha_3~\Pi^+_j +  \alpha_4~\Pi^-_i + \alpha_5~\Pi^-_j\right.\nonumber \\
     & + & \left. \alpha_6~\Pi^+_i\Pi^+_j + \alpha_7~\Pi^-_i\Pi^-_j 
      +  \alpha_8~\Pi^+_i\Pi^-_j + \alpha_9~\Pi^-_i\Pi^+_j~, \right]
        \end{eqnarray}
        is equivalent to a Yang-Baxter operator in the $H3,1$ Hietarinta class
        \begin{equation}
            \begin{pmatrix}
                \alpha_1+\alpha_4+\alpha_5+\alpha_7 & 0 & 0 & 0 \\
                0 & \alpha_1+\alpha_3+\alpha_4+\alpha_9 & 0 & 0 \\
                0 & 0 & \alpha_1+\alpha_2+\alpha_5+\alpha_8 & 0 \\
                0 & 0 & 0 & \alpha_1+\alpha_2+\alpha_3+\alpha_6
            \end{pmatrix},
        \end{equation}
        under conjugation by the $Q$ operator
        \begin{equation}
            Q = \begin{pmatrix}
                0 & q_2 \\
                q_3 & 0
            \end{pmatrix}.
        \end{equation}
        This transforms an arbitrary diagonal 2 by 2 matrix as 
        \begin{equation}
            \begin{pmatrix}
                a & 0 \\ 0 & d
            \end{pmatrix} \rightarrow Q\begin{pmatrix}
                a & 0 \\ 0 & d
            \end{pmatrix}Q^{-1} = \begin{pmatrix}
                d & 0 \\ 0 & a
            \end{pmatrix}.
        \end{equation}
        However this rotated diagonal matrix is equivalent to the unrotated one. This implies that the tetrahedron operator in \eqref{eq:tetrafromcase3YB} is equivalent to the one obtained from the Yang-Baxter of Hietarinta's $H3,1$ class, Family 1 \eqref{eq:unitaryHFamily-1}. Alternatively the conjugation of the Yang-Baxter \eqref{eq:YijRemark4.1} by the $Q$ operator
        \begin{equation}
            Q = \begin{pmatrix}
                q_1 & 0 \\ 0 & q_4
            \end{pmatrix},
        \end{equation}
        reduces it to an operator in the $H3,1$ Hietarinta class
        \begin{equation}
            \begin{pmatrix}
                \alpha_1+\alpha_2+\alpha_3+\alpha_6 & 0 & 0 & 0 \\
                0 & \alpha_1+\alpha_2+\alpha_5+\alpha_8 & 0 & 0 \\
                0 & 0 & \alpha_1+\alpha_3+\alpha_4+\alpha_9 & 0 \\
                0 & 0 & 0 & \alpha_1+\alpha_4+\alpha_5+\alpha_7
            \end{pmatrix}.
        \end{equation}
        However since this diagonal $Q$ matrix commutes with the diagonal $C$ matrix, we can conclude that the tetrahedron operator in \eqref{eq:tetrafromcase3YB} falls in Family 1 \eqref{eq:unitaryHFamily-1}. 
\end{enumerate}
\end{remark}
\begin{remark}
    As the Clifford tetrahedron solution in Case 3 \eqref{eq: projsol} is the same as the tetrahedron solution obtained from the Clifford Yang-Baxter operator \eqref{eq:tetrafromcase3YB}, the former also fall in Family 1 \eqref{eq:unitaryHFamily-1}, the unitary family generated by the $H3,1$ Hietarinta class.
\end{remark}

\section{Summary : Inequivalent unitary tetrahedron operators} 
\label{sec:summaryTetra}
Here we collect all the unitary inequivalent tetrahedron operators obtained {\it via} the different methods in the previous sections. We have discussed two methods for constructing such solutions :
\begin{enumerate}
    \item using Clifford algebras Sec.\ref{sec:unitaryCliffordSolutions}. These are adapted from \cite{padmanabhan2024solving}. We obtain a single unitary family of tetrahedron operators using this method.  
    \item by tensoring Yang-Baxter operators with a third operator Sec.\ref{sec:TetraFromYBsolutions}. This ansatz is adapted from \cite{Hietarinta_1993}. This leads to 12 families of unitary tetrahedron operators.
\end{enumerate}
These methods result in a rather large class of unitary tetrahedron operators acting on a local Hilbert space spanned by qubits. However they are certainly not exhaustive and so do not present a classification of constant unitary 8 by 8 tetrahedron operators. We summarise the 13 unitary classes in Table \ref{tab:7unitaryTFamilies}.
\begin{table}[h!]
\begin{center}
\begin{tabular}{|c|c|c|c|}
    \hline
 & \textbf{Tetrahedron Operator} & \textbf{Constraints} & \textbf{Eigenvalues} \\
 \hline 
  &  $\alpha_0~B_iB_jB_k + \alpha_1~A_iA_jB_k + $ & $\{A,B\}=0,~A^2={\mathbb 1}=B^2$ & $\{\pm \left(\alpha _0-\alpha _1-\alpha _2-\alpha _3\right),$\\
 & $\alpha_2~A_iB_jA_k + \alpha_3~B_iA_jA_k,$ & $\alpha_0 \bar{\alpha}_1+\alpha_1 \bar{\alpha}_0=\alpha_3 \bar{\alpha}_2+\alpha_2 \bar{\alpha}_3$ &  $\pm \left(\alpha _1-\alpha _0-\alpha _2-\alpha _3\right),$\\
$1$ & \textbf{and} & $\alpha_0 \bar{\alpha}_2+\alpha_2 \bar{\alpha}_0=\alpha_3 \bar{\alpha}_1+\alpha_1 \bar{\alpha}_3$ & $\pm \left(\alpha _2-\alpha _0-\alpha _1-\alpha _3\right),$\\
 &  $\alpha_0~A_iA_jA_k + \alpha_1~B_iB_jA_k + $ & $\alpha_0 \bar{\alpha}_3+\alpha_3 \bar{\alpha}_0=\alpha_2 \bar{\alpha}_1+\alpha_1 \bar{\alpha}_2$ & $\pm \left(\alpha _3-\alpha _0-\alpha _1-\alpha _2\right)\}$\\
 & $\alpha_2~B_iA_jB_k +  \alpha_3~A_iB_jB_k$ & $\sum_{i=0}^{3}|\alpha_i|^2=1 ,~\alpha_i\in\mathbb {C} ~\forall~ i$& \\
   \hline
 $2$  & $\kappa~\mathcal{Q}~{\tiny\left[\begin{pmatrix}
            1 & 0 & 0 & 0 \\
            0 & p & 0 & 0 \\
            0 & 0 & q & 0 \\
            0 & 0 & 0 & r
        \end{pmatrix}\otimes Z \right]}~\mathcal{Q}^{-1}$&$Q = \begin{pmatrix}
            q_1 & q_2 \\ -\frac{q_1\bar{q_2}}{\bar{q_4}} & q_4
        \end{pmatrix},$~ & $\{\pm 1,\pm p, \pm q, \pm r
        \}$\\& \textbf{and}~&$|p|=|q|=|r|=1,$&\\&$\kappa~\mathcal{Q}~{\tiny\left[Z\otimes\begin{pmatrix}
            1 & 0 & 0 & 0 \\
            0 & p & 0 & 0 \\
            0 & 0 & q & 0 \\
            0 & 0 & 0 & r
        \end{pmatrix} \right]}~\mathcal{Q}^{-1}$& ~~$ p,q,r\in\mathbb {C}$&\\
        \hline
 $3$ & $\kappa~\mathcal{Q}~\left[\tilde{Y}_{H0,2}\otimes Z\right]~\mathcal{Q}^{-1}$&$Q = \begin{pmatrix}
            q_1 & q_2 \\ -\frac{q_1\bar{q_2}}{\bar{q_4}} & q_4
        \end{pmatrix},$~ &$\{\pm 1,\pm 1,\pm\frac{1+ \mathrm{i}}{\sqrt{2}},\pm\frac{1- \mathrm{i}}{\sqrt{2}}\}$\\
        &\textbf{and}&$|q_1|=|q_4|,$&\\& ~$\kappa~\mathcal{Q}~  \left[Z \otimes \tilde{Y}_{H0,2}\right]\mathcal{Q}^{-1}$& ~~$q_1,q_2,q_3,q_4 \in\mathbb {C},~~$&\\
        
\hline
$ 4$  & $\kappa~\mathcal{Q}~ {\tiny \left[\begin{pmatrix}
            0 & 0 & 0 & p \\
            0 & 0 & 1 & 0 \\
            0 & 1 & 0 & 0 \\
            q & 0 & 0 & 0
        \end{pmatrix}\otimes Z \right]}~\mathcal{Q}^{-1}$&$Q = \begin{pmatrix}
            q_1 & q_2 \\ -\frac{q_1\bar{q_2}}{\bar{q_4}} & q_4
        \end{pmatrix},~~$~ &$\{\pm 1,\pm 1,\pm \sqrt{p q},\pm \sqrt{p q}\}$\\& \textbf{and}~&$|q_1|^2=|q||q_4|^2$,~~$|q_4|^2=|p||q_1|^2,$&\\ &$\kappa~\mathcal{Q}~ {\tiny \left[ Z \otimes \begin{pmatrix}
            0 & 0 & 0 & p \\
            0 & 0 & 1 & 0 \\
            0 & 1 & 0 & 0 \\
            q & 0 & 0 & 0
        \end{pmatrix} \right]} \mathcal{Q}^{-1}$&  ~ $q_1,q_2,q_3,q_4,~p,q \in\mathbb {C}$&\\
         & &  & \\
         \hline
         $ 5$ &$\mathcal{Q}{\tiny\left[\begin{pmatrix}
            0 & 0 & 0 & e^{\mathrm{i}\theta_p} \\
            0 & 0 & 1 & 0 \\
            0 & 1 & 0 & 0 \\
            e^{\mathrm{i}\theta_q} & 0 & 0 & 0
        \end{pmatrix}\otimes \begin{pmatrix}
        0 & 1 \\ \pm \frac{\sqrt{e^{\mathrm{i}\theta_q}}}{\sqrt{e^{\mathrm{i}\theta_p}}} & 0 
    \end{pmatrix}\right] }\mathcal{Q}^{-1}$  &$Q = \begin{pmatrix}
            |q_1|e^{\mathrm{i}\theta_1} & q_2 \\ -\bar{q_2}e^{\mathrm{i}(\theta_1+\theta_4)} & |q_1|e^{\mathrm{i}\theta_4}
        \end{pmatrix},$&$\{\pm e^{-\frac{1}{4} \mathrm{i} (\theta_p-\theta_q)},\pm e^{-\frac{1}{4} \mathrm{i} (\theta_p-\theta_q)},$\\
        &\textbf{and}& $ q_1, q_2\in\mathbb {C}, \quad\theta_ 1, \theta_ 4,\theta_p,\theta_q\in\mathbb {R}$&$\pm e^{-\frac{1}{4} \mathrm{i} (\theta_p+3\theta_q)},\pm e^{-\frac{1}{4} \mathrm{i}(\theta_p+3\theta_q)}\}$\\
         &$\mathcal{Q}{\tiny\left[\begin{pmatrix}
        0 & 1 \\ \pm \frac{\sqrt{e^{\mathrm{i}\theta_q}}}{\sqrt{e^{\mathrm{i}\theta_p}}} & 0 
    \end{pmatrix} \otimes \begin{pmatrix}
            0 & 0 & 0 & e^{\mathrm{i}\theta_p} \\
            0 & 0 & 1 & 0 \\
            0 & 1 & 0 & 0 \\
            e^{\mathrm{i}\theta_q} & 0 & 0 & 0
        \end{pmatrix}\right] }\mathcal{Q}^{-1}$  &&\\
   \hline
 $6$  & $\kappa~\mathcal{Q}~  {\tiny\left[\begin{pmatrix}
            0 & 0 & 0 & 1 \\
            0 & 0 & 1 & 0 \\
            0 & 1 & 0 & 0 \\
            1 & 0 & 0 & 0
        \end{pmatrix}\otimes \begin{pmatrix}
        0 & 1 \\  \pm 1 & 0 
    \end{pmatrix}\right]}~\mathcal{Q}^{-1}$&$Q = \begin{pmatrix}
            q_1 & q_2 \\ q_3 & q_4
        \end{pmatrix},~q_i \in\mathbb {C}~\forall~i,$ &$\{\pm 1,\pm 1,\pm 1,\pm 1\}$\\
        &\textbf{and}&$\bar{q_1}q_1 + \bar{q_3}q_3  =  \bar{q_2}q_2 + \bar{q_4}q_4,~$&\\
       &$\kappa~\mathcal{Q}~ {\tiny\left[ \begin{pmatrix}
        0 & 1 \\  \pm 1 & 0 
    \end{pmatrix} \otimes \begin{pmatrix}
            0 & 0 & 0 & 1 \\
            0 & 0 & 1 & 0 \\
            0 & 1 & 0 & 0 \\
            1 & 0 & 0 & 0
        \end{pmatrix}\right]} \mathcal{Q}^{-1}$& $
        \bar{q_2}q_1 \mp  \bar{q_1}q_2  =  \bar{q_3}q_4 \mp \bar{q_4}q_3$ ~&\\
   \hline
 $7$  & $\kappa~\mathcal{Q}~{ \left[P\otimes M \right]}~\mathcal{Q}^{-1}$& $Q=\begin{pmatrix}
        q_1 & q_2 \\  q_3 & q_4 
    \end{pmatrix}\in U(2)$~ &$\{X^{\pm}_1,X^{\pm}_1,X^{\pm}_1,X^{\pm}_2\}$\\& \textbf{and}&$\&$&\\&~$\kappa~\mathcal{Q}~  {\left[M \otimes P\right]} \mathcal{Q}^{-1}$&$\tiny M=\begin{pmatrix}
        m_1 & m_2 \\  m_3 & m_4 
    \end{pmatrix}\in U(2)$&\\      \hline
\end{tabular}
\end{center}
\caption{The 13 unitary families of tetrahedron operators. The two choices in the first row are equivalent to each other by a local invertible operator. This local operator maps $A$ to $B$. The eigenvalues for the first family have absolute value 1. Note that they are subject to the constraints in the second column and hence are not entirely independent. Each of the 6 entries from Rows 2-7 contain two unitary families each, accounting for the remaining 12 unitary families. Also $ \mathcal{Q} = Q \otimes Q \otimes Q $, and $ \tiny X_1^{\pm} = \frac{1}{2} \left(\pm \sqrt{(m_1-m_4)^2 + 4m_2 m_3} + m_1 + m_4 \right) , X_2^{\pm} = \frac{1}{2} \left(\pm \sqrt{(m_1-m_4)^2 + 4m_2 m_3} - m_1 - m_4 \right) $.}
\label{tab:7unitaryTFamilies}
\end{table}

\begin{remark} The similarity between the spectra of entry 1 and entry 2 from Table \ref{tab:7unitaryTFamilies} gives the impression that they are equivalent to each other. Indeed the case 1 tetrahedron operator can be diagonalized to the case 2 tetrahedron operator
$\begin{tiny}
\begin{pmatrix}
            l & 0 & 0 & 0 \\
            0 & p & 0 & 0 \\
            0 & 0 & q & 0 \\
            0 & 0 & 0 & r
\end{pmatrix}
\end{tiny}\otimes Z$ 
with the identifications 
\begin{eqnarray}
l&=& \alpha _1+\alpha _2+\alpha _3-\alpha _0,\nonumber\\p&=& \alpha _0+\alpha _2+\alpha _3-\alpha _1,\nonumber\nonumber\\q&=& \alpha _0+\alpha _1+\alpha _3-\alpha _2,\nonumber\\r&=& \alpha _0+\alpha _1+\alpha _2-\alpha _3.
\end{eqnarray}
However, as can be checked easily, the matrix $U$ that carries out the diagonalization is not a local gauge transformation, i.e. $U$ cannot be written as $Q\otimes Q\otimes Q$.
\end{remark}

\section{Discussion : Quantum gates}
\label{sec:discussion}
We conclude with an application of unitary tetrahedron operators : realization of universal quantum gate sets. Consider a generic quantum circuit written as
\begin{eqnarray}
    {\cal U}=U_N\cdots U_2 U_1,
\end{eqnarray}
where $U_i$'s are the individual quantum gates. A given set of quantum gates, able to approximate arbitrary quantum operation, within a specified margin of error, is called a universal gate set. For example, all single-qubit gates together with the Controlled NOT (CNOT) gate form a universal set \cite{PhysRevA.52.3457}.  In particular, a universal set must contain gates which can create superposition and entanglement\footnote{These gates also help to explore the connection between topological entanglement (higher dimensional knots) and quantum entanglement. For example the unitary solutions of the constant $2$-simplex (Yang-Baxter) equation help in achieving this for standard knots and links \cite{kauffman2002quantum, kauffman2003entanglement, kauffman2004braiding,Rowell2007ExtraspecialTG,Padmanabhan_2020,Padmanabhan_2021,Quinta_2018}. 
This is done for the spectral parameter dependent Yang-Baxter and its generalizations is in \cite{zhang2005universal, zhang2005yang,Padmanabhan2019QuantumES}.} in order to generate (approximately) any quantum operation. We will show that a significant number of universal gates can be obtained starting from the solutions of the tetrahedron equation. However, there is a possibility that certain gates do not fulfill the tetrahedron equation on their own, but can be derived by multiplying specific unitary solutions of the tetrahedron equation. In other words, they can be represented by quantum circuits, whose component quantum gates solve the tetrahedron equation. All the gates are obtained from the Case 3 of Clifford tetrahedron operators \eqref{eq: projsol} which are equivalent to Family 1 \eqref{eq:unitaryHFamily-1} of the tetrahedron operators obtained from Hietarinta's constant Yang-Baxter solutions. 

\subsection{Single qubit gates}
Consider the unitary tetrahedron operator 
\begin{eqnarray}
    R_{ijk}={\mathbb 1}-2 \Gamma^-_i,
\end{eqnarray}
obtained from \eqref{eq: projsol} by an appropriate choice of coefficients. When $$\Gamma^-=\frac{1}{2}\left(\mathbb{1}-S\right)~,~ \text{where}~S\in\{X,Y,Z\},$$  this yields the corresponding single-qubit gates $R_{ijk}=X_i,Y_i,~ \text{and}~ Z_i,$ respectively. In a similar manner, we also can construct $R_{ijk}=H_i$ by taking $\Gamma^-=\frac{1}{2}\left(\mathbb{1}-H\right)$, where $H$ is the Hadamard matrix $H=\frac{1}{\sqrt{2}}(X+Z)$. More generally, for any single qubit gate $U$, with eigenvalues $\{\lambda_+,\lambda_-\}$, we have the tetrahedron solution
\begin{eqnarray}
    R_{ijk}=U_i=\lambda_+{\mathbb 1}+\left(\lambda_--\lambda_+\right)\Gamma_i^-=\lambda_-{\mathbb 1}+\left(\lambda_+-\lambda_-\right)\Gamma_i^+,
\end{eqnarray}
with $\Gamma^{\pm}=\frac{1}{\lambda_{\pm}-\lambda_{\mp}}\left(U-\lambda_{\mp}{\mathbb 1}\right)$. Therefore essentially all the single qubit gates can be obtained from the tetrahedron solutions. 

\subsection{Two qubit Gates}
Now we will discuss two-qubit gates that can be universal by themselves \cite{PhysRevA.51.1015}. The CNOT gate
\begin{eqnarray}
    \text{CNOT}=
    \begin{pmatrix}
        1 & 0 & 0 & 0\\
        0 & 1 & 0 & 0\\
        0 & 0 & 0 & 1\\
        0 & 0 & 1 & 0
    \end{pmatrix}=\mathbb{1}-\frac{1}{2} \left(\mathbb{1}-Z\right)\otimes \left(\mathbb{1}-X\right),
\end{eqnarray}
is not a tetrahedron operator. Nevertheless, we can represent the CNOT gate as a quantum circuit constructed purely from tetrahedron gates. Consider the two solutions
\begin{eqnarray}
    {\cal R}_{ijk}={\mathbb 1}-2{\Pi}^-_i{\Pi}^-_j,\quad R_{ijk}={\mathbb 1}-2\Gamma^-_j,
\end{eqnarray}
with ${\Pi}^-=\frac{{\mathbb 1}-Z}{2},~ \Gamma^-=\frac{{\mathbb 1}-H}{2}$. It is now easy to obtain the CNOT gate as
\begin{eqnarray}
    {\text CNOT}_{ij;k}=R_{ijk}{\cal R}_{ijk}R_{ijk}.
\end{eqnarray}
The subscript in ${\text CNOT}_{ij;k}$ implies that it acts as the controlled-NOT (CNOT) gate on the first two indices $i,j$ and trivially on the $k$-th qubit. This shows that the universal quantum gate set $\{\text{CNOT, all single-qubit gates}\}$ is obtainable from unitary solutions of the tetrahedron equation.

Furthermore, consider a general operator
\begin{eqnarray}
    {\cal R}_{ijk}(\phi,\psi)=\mathbb{1}-\frac{1}{2} \left(\mathbb{1}-Z\right)\otimes \left(\mathbb{1}-{\cal G}(\phi,\psi)Z\right),
\end{eqnarray}
with ${\cal G}(\phi,\psi)={\rm diagonal}\left[e^{{\rm i}(\phi+\psi)},e^{{\rm i}(\psi -\phi +\pi)}\right]$. It is easy to demonstrate that this is unitary and solves the constant tetrahedron equation. With the same choice of $R_{ijk}=H_j$, we find that
\begin{eqnarray}
    {\sf R}(\phi,\psi)_{ijk}=R_{ijk}{\cal R}(\phi,\psi)_{ijk}R_{ijk}=\left(
\begin{array}{cccc}
 1 & 0 & 0 & 0 \\
 0 & 1 & 0 & 0 \\
 0 & 0 & e^{i \psi } \cos (\phi ) & i e^{i \psi } \sin (\phi ) \\
 0 & 0 & i e^{i \psi } \sin (\phi ) & e^{i \psi } \cos (\phi ) \\
\end{array}
\right).
\end{eqnarray}
Let
\begin{eqnarray}
U_{\phi,\psi}=e^{\mathrm{i} \psi }\left(
\begin{array}{cc}
\cos (\phi ) & \mathrm{i} \sin (\phi ) \\
\mathrm{i} \sin (\phi ) & \cos (\phi ) \\
\end{array}
\right).
\end{eqnarray}
Here, we have
${\sf R}(\phi,\psi)= \Lambda_1\left(U_{\phi,\psi}\right)=\begin{pmatrix}
\mathbb{1}_2 & 0\\
0 & U{\phi,\psi}\\
\end{pmatrix}$, which is commonly referred to as a two-qubit controlled $U_{\phi,\psi}$ gate \cite{PhysRevA.52.3457} and reduces to the CNOT gate for $\phi=\frac{\pi}{2}=-\psi$. A few examples of two-qubit gates are listed in Table \ref{tab:2qubitGates}.
\begin{table}[h!]
\begin{center}
\begin{tabular}{|c|c|c|}
\hline
\textbf{Two-qubit gates}  & \textbf{Tetrahedron operator} &\textbf{ Quantum circuit} \\
\hline
\textit{Controlled-Z gate}&${\mathbb 1}-2{\Pi}^-_i{\Pi}^-_j$ & $\Qcircuit @C=1.2em @R=.8em {
    & \ctrl{1} & \qw \\
    & \gate{Z}  & \qw \\
     & \qw  & \qw \\
     &&\\
}$ \\
\hline
~\textit{CNOT gate} & $\left({\mathbb 1}-2\Gamma^-_j\right)\left[{\mathbb 1}-2{\Pi}^-_i{\Pi}^-_j\right]\left({\mathbb 1}-2\Gamma^-_j\right)$ & $\Qcircuit @C=1em @R=.7em {
    & \ctrl{1} & \qw \\
    & \targ & \qw \\
     & \qw  & \qw \\
     &&\\
}$\\
\hline
~\textit{controlled-$U$ gate }& $H_j\left[{\sf R}(\phi,\psi)_{ijk}\right]H_j$ &$\Qcircuit @C=1em @R=.7em {
    & \ctrl{1} & \qw \\
    & \gate{U_{\phi,\psi}} & \qw \\
      & \qw  & \qw \\
     &&\\
}$ \\
\hline
~\textit{SWAP gate
 }& $P_{ij}$ &$\Qcircuit @C=1em @R=.7em {
    & \qswap &\qw \\
    & \qswap \qwx &\qw \\
      & \qw  & \qw \\
     &&\\
}$  \\
\hline
~\textit{iSWAP gate
 }& $P_{ij}\left[{\mathbb 1}+(-1+\mathrm{i}){\Pi}^-_i+(-1+\mathrm{i}){\Pi}^-_j+(2-2\mathrm{i}){\Pi}^-_i{\Pi}^-_j\right]$ &$\Qcircuit @C=1em @R=.7em {
    & \qswap &\ctrl{1} & \gate{S}&\qw \\
    & \qswap \qwx &\gate{Z} & \gate{S}&\qw \\
      & \qw &\qw & \qw&\qw \\
       &  & & & \\
}$  \\
\hline
\end{tabular}
\end{center}
\caption{Two-qubit gates expressed using tetrahedron operators, assuming an identity operation on the third qubit. $P$ denotes the permutation operators.}
\label{tab:2qubitGates}
\end{table}
\subsection{Three qubit gates}
The CCNOT gate, also referred to as the Toffoli gate, is a crucial three-qubit gate and can be implemented as
\begin{eqnarray}
    T_{ijk}&=&R_{ijk}\left({\mathbb 1}-2 {\Pi}^-_i{\Pi}^-_j{\Pi}^-_k\right)R_{ijk}\nonumber\\
    &=&R_{ijk}\left[{\mathbb 1}-\frac{1}{4}\left({\mathbb 1}-Z_i\right)\left({\mathbb 1}-Z_j\right)\left({\mathbb 1}-Z_k\right)\right]R_{ijk},
\end{eqnarray}
with $R_{ijk}=H_k$. This is also found in \cite{sinha2024toffoli}.
Furthermore, extending the analysis of two-qubit systems straightforwardly leads us to
\begin{eqnarray}
    T(\phi,\psi)=\Lambda_2(U_{\phi,\psi})=\begin{pmatrix}
        {\mathbb 1}_6 & 0\\
        0 & U_{\phi,\psi}\\
    \end{pmatrix}
\end{eqnarray}
with $T(\phi,\psi)_{ijk}=R_{ijk}\left[{\mathbb 1}-\frac{1}{4}\left({\mathbb 1}-Z_i\right)\left({\mathbb 1}-Z_j\right)\left({\mathbb 1}-{\cal G}(\phi,\psi)_k Z_k\right)\right]R_{ijk}$ and ${\cal G}(\phi,\psi)={\rm diagonal}\left[e^{{\rm i}(\phi+\psi)},e^{{\rm i}(\psi -\phi +\pi)}\right]$. Interestingly, this result identifies with the Deutsch gates\cite{deutsch1989quantum}
$$U_{\lambda}=\left(
\begin{array}{ccc}
 {\mathbb 1}_6 & 0 & 0 \\
 0 & \mathrm{i}\cos \lambda  &  \sin \lambda  \\
 0 &  \sin \lambda  & \mathrm{i} \cos \lambda  \\
\end{array}
\right),$$ 
with $\psi=\frac{\pi}{2},\phi=-\lambda$. For completeness, we have included several examples of three-qubit gates in the table\ref{tab:3qubitGates}.

\begin{table}[h!]
\begin{center}
\begin{tabular}{|c|c|c|}
\hline
\textbf{Three-qubit gates } & \textbf{Tetrahedron operator} & \textbf{Quantum circuit} \\
\hline
~\textit{CCZ gate} & $(CCZ)_{ijk}=\left[{\mathbb 1}-2{\Pi}^-_i{\Pi}^-_j{\Pi}^-_k\right]$ & $\Qcircuit @C=1.5em @R=.6em {
& \ctrl{2} & \qw \\
& \ctrl{-1} &  \qw \\
& \gate{Z} & \qw \\
}$\\
\hline
~\textit{Toffoli gate }& $T_{ijk}=H_k\left[{\mathbb 1}-2{\Pi}^-_i{\Pi}^-_j{\Pi}^-_k\right]H_k$ &$\Qcircuit @C=1.5em @R=.6em {
& \ctrl{2} & \qw \\
& \ctrl{-1} &  \qw \\
& \targ & \qw \\
}$ \\
\hline
\textit{Deutsch gate} &$ H_k\left[{\mathbb 1}-{\Pi}^-_i{\Pi}^-_j\left({\mathbb 1}-{\cal G}(-\lambda,\frac{\pi}{2})_k Z_k\right)\right]H_k$ & $\Qcircuit @C=1.5em @R=.6em {
& \ctrl{2} & \qw \\
& \ctrl{-1} &  \qw \\
& \gate{ U_{(-\lambda,\frac{\pi}{2})}}&  \qw \\ }$
 \\
\hline
~\textit{Margolus gate} & $\left[{\mathbb 1}-2 {\Pi}^-_i{\Pi}^+_j{\Pi}^-_k\right]T_{ijk}$ & $\Qcircuit @C=1.em @R=.3em {
&\qw & \ctrl{2} &\qw& \ctrl{2}&\qw \\
&\gate{X}& \ctrl{-1}&\gate{X} &  \ctrl{-1} &\qw\\
&\qw& \gate{Z} &\qw& \targ&\qw \\
}$\\
\hline
~\textit{Fredkin gate(CSwap)} & $H_j\left[(CCZ)_{ijk}\right]H_j \left[T_{ijk}\right]H_j\left[(CCZ)_{ijk}\right]H_j$ &$\Qcircuit @C=.2em @R=.5em {
&\qw & \ctrl{2} &\qw& \ctrl{2}&\qw &\ctrl{2}&\qw&\qw\\
&\gate{H}& \ctrl{-1}&\gate{H} &  \ctrl{-1} &\gate{H}&\ctrl{-1}&\gate{H}&\qw\\
&\qw& \gate{Z} &\qw& \targ&\qw &\gate{Z} &\qw&\qw\\
}$\\
 \hline
\end{tabular}
\end{center}
\caption{Three qubit gates written as products of tetrahedron operators. The Margolus gate \cite{song2003simplified} is a simplified Toffoli gate, differing only by a relative phase.}
\label{tab:3qubitGates}
\end{table}
\begin{remark}
    The applications of Hietarinta unitary solutions $(Y \otimes M)$ are connected to constructing quantum circuits that utilize matchgates\cite{VALIANT2002}. A matchgate is any unitary operation that can be represented by the following matrix:
\begin{eqnarray*}
 U_M &=& \left(
\begin{array}{cccc}
p_1 & 0 & 0 & p_2 \\
0 & s_{1} & s_{2} & 0 \\
0 & s_{3} & s_{4} & 0 \\
p_3 & 0 & 0 & p_{4} \\
\end{array}
\right), ~~  P = \left(
\begin{array}{cc}
p_1 & p_2 \\
 p_3 & p_4\\
\end{array}
\right), ~~S =\left( \begin{array}{cc}
s_1 & s_2 \\
 s_3 & s_4\\
\end{array}\right),
\end{eqnarray*}
where $ \{P,S\}\in U(2)~ \text{or}~ SU(2)$ with the same determinant. In our exploration, it is evident that the five families of $4\times 4$ constant Yang-Baxter operators correspond to matchgates. This opens up intriguing possibilities for the effectiveness of quantum circuits based on matchgates. For further insights, readers can refer to \cite{zhang2024geometricrepresentationsbraidyangbaxter, jozsa2008matchgates, ermakov2024unified}. We look forward to explore into these aspects in our future work.
\end{remark}

\section*{Acknowledgments}
VK is funded by the U.S. Department of Energy, Office of Science, National Quantum Information Science Research Centers, Co-Design Center for Quantum Advantage ($C^2QA$) under Contract No. DE-SC0012704. $C^2QA$ (led the research through discussions and idea generations, and participated in preparing and editing the draft) in this research. The work of VKS is supported by ``Tamkeen under the NYU Abu Dhabi Research Institute grant CG008 and  ASPIRE Abu Dhabi under Project AARE20-336''. VKS and PP thank the organizers of the conference ``Quantum Information and Matter'', May 27-May 31, 2024 at NYU Abu Dhabi, where part of this work was developed. 

\appendix

\section{Examples of constructing higher simplex from lower simplex solutions}
\label{app:lower2higher}
We will now elaborate more on the ways to obtain higher simplex operators from lower simplex solutions. This can be seen as a generalization of the tetrahedron ansatz used in \cite{Hietarinta_1993}. A more thorough analysis is reserved for a future work.

Consider the vertex form of the 4-simplex equation 
\begin{eqnarray}
    & R_{1234}R_{1567}R_{2589}R_{368,10}R_{479,10} & \nonumber \\
    = & R_{479,10}R_{368,10}R_{2589}R_{1567}R_{1234}. & 
\end{eqnarray}
A careful look at the index structures of this equation reveals the possibility of constructing 4-simplex operators from 2- and 3-simplex operators. For example the indices $1,2$, $1,5$ and $2,5$, appearing in that order on the left hand side and in the reverse order on the right hand side, coincide with a Yang-Baxter equation on the indices $\{1,2,5\}$. In a similar manner the 4-simplex equation also contains Yang-Baxter equations on indices $\{5,6,8\}$ and $\{8,9,10\}$. Thus we expect the following ansatze ($R_{ijkl}$) to solve the 4-simplex equation :
\begin{eqnarray}
   Y_{ij}M_kM_l~;~M_iY_{jk}M_l~;~M_iM_jY_{kl}. 
\end{eqnarray}
This is indeed the case when the $Y$'s are Yang-Baxter operators and when $\left[Y, M\otimes M\right]=0$. This is similar to the condition on the $Y$'s and $M$'s to satisfy the tetrahedron equation. This logic continues to hold for the higher simplex cases starting from the vertex form of the 5-simplex equation
\begin{eqnarray}
    & R_{12345}R_{16789}R_{26,10,11,12}R_{37,10,13,14}R_{48,11,13,15}R_{59,12,14,15} & \nonumber \\
    = & R_{59,12,14,15}R_{48,11,13,15}R_{37,10,13,14}R_{26,10,11,12}R_{16789}R_{12345}. &
\end{eqnarray}
In this case the Yang-Baxter equation appears on the index sets $\{1,2,6\}$; $\{6,7,10\}$; $\{10,11,13\}$ and $\{13,14,15\}$. Thus the ansatze which solve the 5-simplex equation are :
\begin{eqnarray}
Y_{ij}M_kM_lM_m~;~M_iY_{jk}M_lM_m~;~M_iM_jY_{kl}M_m~;~M_iM_jM_kY_{lm}
\end{eqnarray}
when $Y$ is a Yang-Baxter operator and $\left[Y, M\otimes M\right]=0$. The logic to construct other higher simplex operators from Yang-Baxter operators is now evident. For instance we can construct 4-simplex and 5-simplex operators from tetrahedron operators using the ansatze :
\begin{eqnarray}
    T_{ijk}M_l~~;~~T_{ijk}M_lM_m,
\end{eqnarray}
respectively. In this case we require that $\left[T, M\otimes M\otimes M\right]=0$ apart from $T$ satisfying the tetrahedron equation. Further studies on the structure of these solutions will be studied elsewhere.

\section{$A$, $B$, $C$ words and generalised tetrahedron equations}
\label{app:3from2}
We have seen that the tetrahedron operators \eqref{eq:tetraABC-1} and \eqref{eq:tetraABC-2} are two possible ways to obtain 3-simplex operators from 2-simplex solutions. This is done by including an additional operator $C$ that anticommutes with $A$ and $B$. Now we look at other possible words on the tensor product space $V\otimes V\otimes V$ built using $A$, $B$ and $C$. Consider 
\begin{equation}
    R_{ijk} = \alpha~A_iC_jA_k + \beta~B_iC_jB_k~;~\alpha, \beta\in\mathbb{C}.
\end{equation}
This operator solves a `tetrahedron-like' equation given by
\begin{equation}
    R_{123}R_{145}R_{246}R_{356} = R_{356}R_{246}^{(-)}R_{145}^{(-)}R_{123},
\end{equation}
where $R^{(-)}_{ijk}$ is given by,
\begin{equation}
    R^{(-)}_{ijk} = \alpha~A_iC_jA_k - \beta~B_iC_jB_k.
\end{equation}
This equation partially resembles the anti-3-simplex equation, which was introduced and discussed in \cite{padmanabhan2024solving}.
Other choices for words in $A$, $B$ and $C$ are interpreted as 3-simplex operators built from anti-Yang-Baxter solutions. These operators satisfy tetrahedron-like equations that are similar in structure to the anti-3-simplex equation. Such $R_{ijk}$'s take the form
\begin{eqnarray}
    & \alpha~A_iB_jC_k + \beta~B_iA_jC_k, & \\
    & \alpha~A_iC_jB_k + \beta~B_iC_jA_k,& \\
    & \alpha~C_iA_jB_k + \beta~C_iB_jA_k. & 
\end{eqnarray}
These satisfy variants of the anti-3-simplex equations :
\begin{eqnarray}
    R_{123}R_{145}R_{246}R_{356} & = & -R_{356}R_{246}R_{145}R_{123}, \\
    R_{123}R_{145}R_{246}R_{356} & = & -R_{356}R_{246}^{(-)}R_{145}^{(-)}R_{123}, \\
    R_{123}R_{145}R_{246}R_{356} & = & -R_{356}R_{246}R_{145}R_{123}. 
\end{eqnarray}

\paragraph{More examples :} Now we present a few more examples to illustrate the construction of higher simplex operators from lower simplex solutions. We continue to use three mutually anticommuting operators in the examples to follow. We start with 4-simplex solutions built using 2-simplex operators. Two choices for $R_{ijkl}$ are
\begin{eqnarray}
    & \alpha~A_iA_jC_kC_l + \beta~B_iB_jC_kC_l, & \\
     & \alpha~C_iC_jA_kA_l + \beta~C_iC_jB_kB_l. &
\end{eqnarray}
An example of an anti-4-simplex operator \cite{padmanabhan2024solving} is obtained from an anti-3-simplex solution appended with a third operator $C$
\begin{equation}
    R_{ijkl} = A_iA_jA_kC_l + B_iB_jB_kC_l.
\end{equation}
This satisfies the anti-4-simplex equation,
\begin{eqnarray}
    & R_{1234}R_{1567}R_{2589}R_{368,10}R_{479,10} & \nonumber \\
    & = -R_{479,10}R_{368,10}R_{2589}R_{1567}R_{1234}.
\end{eqnarray}
We close with an example of a 5-simplex operator built using a 4-simplex solution,
\begin{equation}
    R_{ijklm} = A_iA_jA_kA_lC_m + B_iB_jB_kB_lC_m.
\end{equation}

\section{Spectral parameter dependent tetrahedron operators : `Baxterization' of tetrahedron operators}
\label{app:BaxterizationT}
The constant Yang-Baxter operator can be made to depend on spectral parameters through the process of Baxterization \cite{Jones1990}. Following this the new $R$-matrix can be used to construct integrable quantum systems and then solve them using the algebraic Bethe ansatz. As discussed in the Introduction (Sec. \ref{sec:Introduction}), these operators can also be used to simulate quantum systems using integrable quantum circuits. It should also be noted that the spectral parameter dependent version is not the same as the constant Yang-Baxter operator, in fact it lives in the braid group algebra generated by the latter. We will briefly discuss how to carry out a process analogous to Baxterization for the constant tetrahedron operators. 

From the structure of the constant tetrahedron operators in \eqref{eq:tetraYM} and \eqref{eq:tetraMY}, it is clear that we can achieve a `partial Baxterization'\footnote{By this we mean that we inherit the spectral parameter dependence from the lower simplex operator, which is the Yang-Baxter operator in this case. Physically this will mean that some of the scattering parameters assume fixed values.} by lifting the Baxterized Yang-Baxter operator. This implies that the spectral parameter dependent tetrahedron operator takes the form 
\begin{equation}\label{eq:tetraYM-SP}
    T_{ijk}(\mu_{ijk}) = Y_{ij}(\mu_{ij})M_k.
\end{equation}
For simplicity we choose the single indexed operator $M$ to be independent of the spectral parameter. We suppose that this can be relaxed to a more general dependence on the spectral parameter. However this will require more analysis and we stick to the simplest case here. Now when $Y$ satisfies,
\begin{equation}
    Y_{12}(\mu_{12})Y_{13}(\mu_{13})Y_{23}(\mu_{23}) =  Y_{23}(\mu_{23})Y_{13}(\mu_{13})Y_{12}(\mu_{12}),
\end{equation}
and $\left[M\otimes M, Y(\mu)\right]=0$, the tetrahedron operator in \eqref{eq:tetraYM-SP} satisfies,
\begin{eqnarray}
    & T_{123}(\mu_{123})T_{145}(\mu_{145})T_{246}(\mu_{246})T_{356}(\mu_{356}) & \nonumber \\
    = & T_{356}(\mu_{356})T_{246}(\mu_{246})T_{145}(\mu_{145})T_{123}(\mu_{123}), &
\end{eqnarray}
a spectral parameter dependent tetrahedron equation. A more non-trivial version of the spectral parameter dependent tetrahedron operator will depend on the algebra satisfied by the tetrahedron operators. This is reserved for a future work.

\bibliographystyle{acm}
\normalem
\bibliography{refs}

\end{document}